\documentclass[useAMS,usenatbib]{mnras}

\usepackage{natbib}
\usepackage[english]{babel}
\usepackage[babel]{csquotes}
\usepackage{indentfirst}
\usepackage{setspace}
\usepackage{fancyhdr}
\usepackage{graphicx,subfigure}
\usepackage{booktabs}
\usepackage{amsmath}
\usepackage[format=plain,font=footnotesize,margin={1cm,1cm},labelfont=bf]{caption}
\usepackage{tensor}
\usepackage{color}
\usepackage{placeins}

\def\be{\begin{equation}} 
\def\ee{\end{equation}} 
\def\ba{\begin{eqnarray}} 
\def\ea{\end{eqnarray}}

\def\cc{\,{\rm {cm^{-3}}}}

\def\gsim{\lower.5ex\hbox{\gtsima}} 
\def\lsim{\lower.5ex\hbox{\ltsima}} \def\gtsima{$\; \buildrel > \over 
\sim \;$} \def\ltsima{$\; \buildrel < \over \sim \;$} \def\prosima{$\; 
\buildrel \propto \over \sim \;$} \def\gsim{\lower.5ex\hbox{\gtsima}} 
\def\lsim{\lower.5ex\hbox{\ltsima}} 
\def\simgt{\lower.5ex\hbox{\gtsima}} 
\def\simlt{\lower.5ex\hbox{\ltsima}} 
\def\simpr{\lower.5ex\hbox{\prosima}}   
  
 \def\gtsima{$\; \buildrel > \over \sim \;$} 
\def\ltsima{$\; \buildrel < \over \sim \;$} 
\def\gsim{\lower.5ex\hbox{\gtsima}} 
\def\lsim{\lower.5ex\hbox{\ltsima}} 
\def\simgt{\lower.5ex\hbox{\gtsima}} 
\def\simlt{\lower.5ex\hbox{\ltsima}} 
\def\simpr{\lower.5ex\hbox{\prosima}}

\def\E3{{\cal E}_{\rm g}^{III}}

\def\r12{r_{1/2}} 
\def\x12{x_{1/2}} 
\def\v12{v_{1/2}}

\pagerange{\pageref{firstpage}--\pageref{lastpage}} \pubyear{2016}
\title[Stellar BH growth galaxies]{Growth problems of stellar black holes in early galaxies}
\author[M.C. Orofino, A. Ferrara, \& S. Gallerani]
{M.C.~Orofino$^1$\thanks{Email: carmela.orofino@sns.it},
  A.~Ferrara$^{1,2}$, S. Gallerani$^{1}$\\
$^1$Scuola Normale Superiore, Piazza dei Cavalieri 7, 1-56126 Pisa, Italy\\
$^2$Kavli IPMU, The University of Tokyo, 5-1-5 Kashiwanoha, Kashiwa 277-8583, Japan\\
}
\date{}

\begin{document}
\maketitle

\label{firstpage}

\begin{abstract}
The nature of the seeds of the observed high-$z$ super-massive black holes (SMBH) is unknown. Although different options have been proposed, involving e.g. intermediate mass direct collapse black holes, BH remnants of massive stars remain the most natural explanation. To identify the most favorable conditions (if any) for their rapid growth, we study the accretion rate of a $M_\bullet = 100 M_\odot$ BH formed in a typical $z=10$ galaxy under different conditions (e.g. galaxy  structure, BH initial position and velocity). We model the galaxy baryonic content and follow the BH orbit and accretion history for $300 \, \rm Myr$ (the time span in $10 > z > 7$), assuming the radiation-regulated accretion model by \cite{park_ricotti:2013}. We find that, within the limits of our model, BH seeds cannot grow by more than $30 \%$, suggesting that accretion on light-seed models are inadequate to explain high-$z$ SMBH. We also compute the X-ray emission from such accreting stellar BH population in the $ [0.5 - 8 ]$ keV band and find it comparable to the one produced by high-mass X-ray binaries. 
This study suggests that early BHs, by X-ray pre-heating of the intergalactic medium at cosmic dawn, might leave a specific signature on the HI 21 cm line power spectrum potentially detectable with SKA.
\end{abstract}


\section{Introduction}
\label{introduction}
The presence of extremely bright quasars revealed by high-redshift observations \citep{fan:2006, willott:2007, mortlock:2011, banados:2014, venemans:2015, jiang:2015, wu:2015} represents a challenge to our understanding of the early universe. As the large bolometric luminosity ($\approx 10^{46} \rm erg \, s^{-1}$) and hardness of their spectra rule out a stellar nature of their central engine, quasars are now canonically interpreted as being powered by super-massive black holes (SMBH). In spite of this progresses, additional questions about their origin are nevertheless left unanswered.
As an example, assembling the mass $M = 2 \times 10^9 M_\odot$ deduced for {the quasar ULAS J1120+0641 at $z = 7.085$ (i.e. 770 billion years after the Big Bang) \citep{mortlock:2011}} requires the SMBH seed to sprout with mass larger than $400 M_\odot$ and accrete at least at the Eddington rate throughout its lifetime. Unfortunately, both these preconditions are uncomfortably demanding due to current estimates of the mass of first stars (\citealt{omukai:2010, greif:2011, bromm:2011}) and to the radiative feedback effects that limit the matter accretion rate  (\citealt{johnson:2007, alvarez:2009}).

In confident expectation of illuminating observations from present and future telescopes, a wide range of SMBHs formation scenarios has been explored theoretically, each of them providing a different, and potentially testable, explanation of the seeds nature
(see \citealt{volonteri:rev:2010, volonteri:2012, haiman:2013, latif:2016} for reviews on this topic).
Currently, the most popular interpretation relies on the monolithic gravitational collapse of pristine gas in atomic-cooling (virial temperature $T_{vir} \geq 1.5 \times 10^4 \rm K$), $\rm H_2$-free primordial halos, leading to a single $10^{4 - 6} M_\odot$ direct-collapse black hole (DCBH).

Although this process would be ideal to account for the observed rapid growth of the SMBHs (see, e.g. \citealt{rees:1984,ferrara:2014} for a discussion) this scenario is still awaiting a solid observational confirmation. 
The detection of powerful Ly$\alpha$ line emission, combined with the initial claim of a prominent HeII line (recently unconfirmed by the SILVERRUSH project \citep{shibuya:2017}) and absence of detectable metal lines, in the $z=6.6$ galaxy CR7 raised hopes that a DCBH was finally identified.
The peculiar emission properties of CR7 \citep{sobral:2015} could in principle arise from sources formed by pristine gas: either a cluster of Population III stars or a DCBH.
Following \cite{pallottini:2015}, who compared the spectrum expected from a $10^5 M_\odot$ DCBH with the one of CR7, several authors \citep{agarwal:2016,hartwig:2016,visbal:2016,smith:2016,smidt:2016,dijkstra:2016} investigated the possible identification of CR7 with a DCBH.
Although  photometric observations by \cite{bowler:2016} in the optical, near-IR and the mid-IR bands seem to favour a faint-AGN nature of CR7, 
\cite{pacucci:2017} confirm that those observations are still consistent with a DCBH and suggest that future spectroscopic observations with JWST are required to solve the puzzle.
To date, although \cite{pacucci:2016} proposed two good candidates ($z>6$ objects in the CANDELS/GOODS-S field with X-ray counterpart revealed by Chandra), no detection of early SMBH progenitors has been confirmed. 

Alternatively, heavy seeds formation could occur in low-metallicity, dense stellar clusters where, due to the energy equipartition theorem, the most massive members tend to sink to its center.
According to several authors \citep{spitzer:1969,vishniac:1978, begelman:1978,lee:1987,quinlan:1990}, if such mass segregation occurs within a time-scale of $3 \rm \, Myr$, a very massive star can be assembled, eventually leading to the formation of an intermediate mass BH at the end of its evolution. \cite{omukai:2008} and \cite{devecchi:2009} calculated that these dense, low-metallicity clusters could form at $z \sim 15$ in dark matter halos of about $10^8 M_\odot$.
{Moreover, \cite{davies:2011}  suggested that runaway merger of stellar mass BHs in clusters  driven by free-fall inflow of self-gravitating gas can produce a BH seed above $10^5 M_\odot$. \cite{lupi:2014} found that this route is feasible, peaks at $z<10$ and it is independent on the metal content of the parent cluster.
}

Given the above uncertainties on the formation mechanisms, abundance and properties of intermediate-mass seeds, remnants of massive stars remain the simplest, and perhaps most natural, SMBH seed candidates. Their mass depends on the metallicity, mass and rotation speed of the parent star and is ultimately associated to the properties of the birth environment \citep{ciardi:2005}; however, recent results have converged on intermediate masses ($\approx 50 M_\odot$) (\citealt{omukai:2010, greif:2011}), i.e. much lighter than those characterizing DCBHs. Hence, it is interesting to clarify whether favorable conditions for the rapid growth of these light seeds may exist in primordial halos. 

SMBH growth from light seeds has already been preliminary studied by other authors. For example, \cite{alvarez:2009} studied stellar BH accretion including radiative feedback effect in an adaptive mesh refinement cosmological simulation. Due to radiative feedback, their black holes do not gain mass at a sufficiently high 
rate to grow into a fully-fledged SMBH. \cite{lupi:2016} present results from a suite of numerical high-resolution
simulations aimed at studying the growth of stellar-mass BHs at super-Eddington accretion; they found that some 
of their BHs reach $10^3 - 10^4 M_\odot$, making them good seeds candidates. Clearly, due to their computational complexity, numerical simulations can explore only a narrow region of the parameter space; here we propose a semi-numerical approach that allows a much wider investigation on the physical parameter space of the problem.

In this analysis, we focus on the growth of a stellar mass BH with reference mass of $10^2 M_\odot$ in a $z=10$ galaxy.
In particular, we derive its accretion history including radiative feedback effects under different initial conditions (galaxy structure, BH position and velocity) by following its orbit, and predict its time-dependent accretion luminosity from a detailed spectral model\footnote{Throughout the paper,  we assume a flat Universe with the following cosmological parameters:  $\Omega_{\rm m} = 0.308$, $\Omega_{\Lambda} = 1- \Omega_{\rm m} = 0.692$, and $\Omega_{\rm b} = 0.048$,  where $\Omega_{\rm M}$, $\Omega_{\Lambda}$, $\Omega_{\rm b}$ are the total matter, vacuum, and baryonic densities, in units of the critical density,   and  $h$ is the Hubble constant in units of 100 km/s \citep{planck:2016}.}. The paper is organized as follows: in Sec. \ref{sec:model} we present our model (describing the host galaxy and BH accretion/emission), Sec. \ref{sec:results} and \ref{sec:emission} contain the results. Conclusions are given in  Sec. \ref{sec:conclusions}.

\section{Model}
\label{sec:model}
Our main aim is to model the accretion and growth history of a stellar mass black hole orbiting in a high-redshift galaxy ($7<z<10$). 
In the following, we describe how we set up the galaxy density within the host dark matter halo. Next, we describe how we compute the black hole accretion, associated emission and dynamics.   
\subsection{Host galaxy structure}
\label{ssec:host_gal}
The black hole accretion rate (i.e. the mass accreted per unit time) strongly depends on the density of the surrounding material.
Hence, accretion models have to define the gas distribution in the host galaxy. Our analysis focuses on spherical and disk galaxies, hosted in $z=10$ dark matter (DM) halos. The structural relations among the virial halo mass ($M_{vir}$), the virial temperature ($T_{vir}$) and radius ($r_{vir}$) at $z=10$ are as follows:
\begin{eqnarray}
M_{vir} = 0.54 \times 10^8 h^{-1} \left(\frac{T_{vir}}{19800 {\rm K }}\right)^{3/2} M_\odot \\
r_{vir} = 1.05 \; h^{-2/3}\left( \frac{M_{vir}}{10^8 M_\odot}\right)^{1/3}  \rm \, kpc,
\label{virial}
\end{eqnarray} 
where we have assumed a gas mean molecular weight $\mu = 1.2$ appropriate for a neutral, primordial gas. We assume that, { within $ r_{vir}$}, virialized dark matter halos have an universal (spherically averaged) density profile, according to the numerical simulations of \cite{navarro:1995} (NFW, hereafter):
\begin{equation}
\rho_{NFW}(r) = \frac{\rho_c \delta_c}{cx(1+cx)^2},
\label{eq:NFW}
\end{equation}
where $x=r / r_{vir}$, $\rho_c = 3 H^2/8 \pi G$ is the critical density, and $c$ is the halo concentration parameter taken from \cite{prada:2012}; $\delta_c = 200 c^3/3 F(c)$ is a characteristic overdensity and
\begin{equation}
F(c) = \ln(1+c) - \frac{c}{1+c}.
\label{eq:F(c)}
\end{equation}
{ Outside the virial radius, a cut-off on the mass is assumed.}
The values of $M_{vir}$, $r_{vir}$ and $c$  for the reference $T_{vir} = (0.5,1,5,10)\times 10^4 \rm K$ DM halos are given in 
Tab. \ref{tab:halos}. 
{ According to \cite{dimatteo:2017} massive BHs (up to $10^8 M_\odot$) at $z=8$ are hosted in halos with virial temperature above $10^6 \, \rm K$, however these haloes are far less abundant then the $10^{4-5} \, \rm K$ haloes considered in this work.
}
\begin{table}
		\centering
		\begin{tabular}{c|c|c|c}
			\hline
			$T_{vir} [10^4 {\rm K}]$ & $M_{vir} [10^7 {M_\odot}]$ & $r_{vir} [\rm kpc]$ & $c$ \\
			\hline
			$0.5$ & $1.0$ & $0.6$ & $4.3$\\
			$1$ & $2.9$ & $0.9$ & $4.1$\\
			$5$ & $32.7$ & $2.0$ & $4.0$\\
			$10$ & $92.4$ & $2.9$ & $3.9$\\
			\hline
		\end{tabular}

\caption{Parameters of dark matter halos considered in this work.}
\label{tab:halos}
\end{table}	
The halo mass contained within a radius $r$, 
\begin{equation}
M(r) = \int^r_0 4 \pi {r'}^2 \rho(r') dr'= M_{vir} \frac{F(cx)}{F(c)},
\label{eq:M_NFW}
\end{equation}
can be translated into a circular velocity
\begin{equation}
v_c^2(r) = \frac{G M(r)}{r} = v_c^2 \frac{F(cx)}{xF(c)},
\label{eq:v_c}
\end{equation}
where $v_c^2 = G M_{vir}/r_{vir}$ . The escape velocity from radius $r$ is\footnote{ The approximation in Eq. (7) comes from the integral up to infinity. This approximation slightly overestimates the value of the
escape velocity, but the density profile Eq. (8) is correct within 1\%.}
{ 
\begin{equation}
v_e^2(r)= 2 \int^{r_{vir}}_r \frac{G M(r')}{r'^2}dr' \approx 2 v_c^2 \frac{F(cx) + \frac{cx}{1+cx}}{xF(c)}
\label{eq:v_e}
\end{equation}
}and reaches a maximum  $2 v_c^2 [c/F(c)]$ at the halo center.

\begin{figure}
                \centering
                \includegraphics[width=0.45\textwidth]{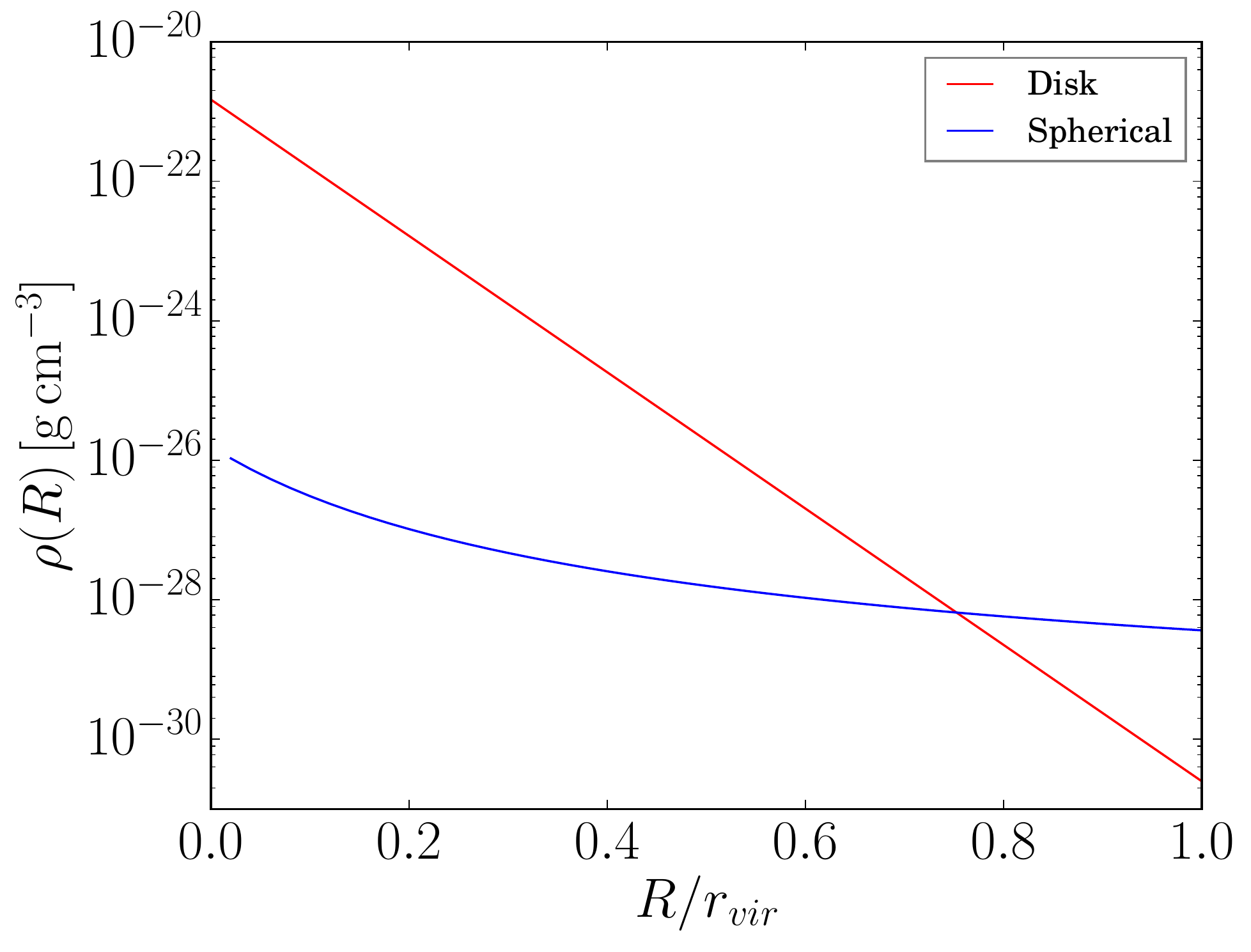}
                \caption{Gas density profile in a $T_{vir} = 10^4 \rm K$ DM halo for the spherical (blue line, eq. \ref{eq:makino}), and disk-like galaxy model (red line). Note that the disk central density is about $10^5$ times higher than the spherical case.}
\label{fig:rho_baryons}
\end{figure}

After virialization, the gas attains an isothermal equation of state (i.e. $P \propto n^\gamma$, with adiabatic index $\gamma = 1$). This entails a sound speed $c_s = (\gamma k_B T_{vir}/ \mu m_p)^{1/2} =  8.3 T_{vir,4}^{1/2} \,\rm km \, s^{-1}$, where $m_p$ is the proton mass\footnote{Throughout the paper we use the notation $Y_x = Y/10^x$}.
The gas settles in approximate hydrostatic equilibrium in the dark matter potential, and its density profile is \citep{makino:1998}
\begin{equation}
\rho_g = \rho_0 \exp\left\{-\frac{\mu m_p}{2 k T_{vir}} \left[ v_e^2(0) - v_e^2(r)\right]\right\};
\label{eq:makino}
\end{equation}
the central density $\rho_0 = 840 h^{-2} \rho_c$ has been obtained by imposing that the gas mass fraction within $r_{vir}$ is equal to $\Omega_b/\Omega_m$. The gas density profile $\rho_g(r)$ in a DM halo with $T_{vir} = 10^4 \rm K$ is shown in Fig.  \ref{fig:rho_baryons}.

The dissipational collapse of a fraction of the baryonic component (whose specific angular momentum is assumed to be the same as the DM one, $J_{vir}$) leads to the formation of a disk. In an almost neutral, primordial composition\footnote{Heavy element and dust cooling become important only for metallicities $Z \simgt 10^{-4}$.} gas, cooling is essentially provided by $\rm H_2$ rotational lines,
which are excited above 300 K. However, $\rm H_2$ molecules can be dissociated by an intense Lyman-Werner background \citep{bromm:2003}; in this case, cooling process is mainly provided by the Ly$\alpha$ emission of neutral H. Such process works above some minimum temperature $\approx 8000 \, \rm K$. In this work, we assume that the disk can cool down to $T_d = 300 \, \rm K$ due to H$_2$ cooling. 
\footnote{ The sound speed in the disk is: $ c_{s,d} (T_d = 300 \, {\rm K}) = 1.5 \rm \, km \, s^{-1}$.}
We refer to \cite{omukai:2005} for a detailed description of the cooling process.

Following \cite{mo:1998a} and \cite{mo:1998b},
let us assume the disk settles onto an exponential surface density profile
\begin{equation}
\Sigma (R) = \Sigma_0 e^{-R/R_d},
\label{eq:surf_density}
\end{equation}
where $R = \sqrt{x^2+y^2}$ is the radial coordinate on the disk plane, $R_d$ the scale length, and
$M_d = 2 \pi \Sigma_0 R_d^2$ the total mass of this disk (taken to be 5\% of the total mass, see below).
We further assume that baryons not ending up in the disk retain the hydrostatic distribution (eq. \ref{eq:makino}). 
The disk gravity induces a contraction of the DM, whose modeling is particularly challenging. It is also possible that violent baryonic processes occurring in the galactic disk (e.g. SN feedback) flatten the central DM density cusp into a constant density core \citep{pontzen:2012}. However, since the disk mass is $5 \%$ of the system total mass, we neglect in this work any variation arising
from disk formation on the NFW profile. 
{ 
At $T_d = 300 \rm \, K$, the Toomre stability criterium predicts that the galactic disk would fragment into clumps. 
In this work, the BH travels in a uniform medium, even if a more rigorous approach should account for the probability to pass through clumps: nevertheless, we calculated that if the BH spends all the integration time in a clumpy medium { ($n \approx 10^2 \, \rm cm^{-3}$, \cite{pallottini:2017}) its maximum mass increment would be about $30\%$, in agreement with the main result of this study.}
\cite{lupi:2016} state that the large reservoir of dense cold gas in a clumpy medium allows super-Eddington accretion via slim-disk evolution.
\cite{fiacconi:2013}, \cite{tamburello:2017}, \cite{souza:2017} studied the orbital decay of massive BH pairs in a clumpy circumnuclear disk, important to predict BH dynamics in galaxy mergers remnants.
Indeed, our study neglects that the density field of galaxies can be affected by galaxy mergers; for a comprehensive review of the evolution of galaxy structure from first galaxies to the local universe see \cite{conselice:2014}.
}

The disk properties can be derived in a simple way from those of the halo. The halo rotation can be expressed through the spin parameter,
\begin{equation}
\lambda = \frac{J_{vir} \sqrt{|E|}}{GM_{vir}^{5/2}},
\label{eq:lambda}
\end{equation}
which is a measure of the rotational-to-gravitational energy of the system. Numerical simulations show that $\lambda$ is basically independent of halo mass, redshift and cosmology, and that its distribution for cold dark matter halos is well fitted by a lognormal distribution that peaks at $\lambda = 0.05$ (\citealt{barnes:1987, maccio:2007}).
We assume that the disk mass and angular momentum are a fixed fraction of the halo ones, so that:
\begin{equation}
m_d = \frac{M_d}{M_{vir}} ,  \;
j_d = \frac{J_d}{J_{vir}}.
\label{eq:md,jd}
\end{equation}
According to \cite{mo:1998a}, consistency with observational results requires $j_d \approx m_d \approx 0.05$.

For a given rotation curve $v_c(r)$, the angular momentum of the disk is:
\begin{equation}
J_d = \int^{r_{vir}}_0 2 \pi \Sigma(r) r^2 v_c(r) dr = 2 M_d v_c R_d F_R,
\label{eq:J_d}
\end{equation}
{
with
\begin{equation}
F_R = \frac{1}{2} \int^{{r_{vir}}/{R_d}}_0 u^2 e^{-u} \frac{v_c(u R_d)}{v_c} du, 
\label{eq:F_R}
\end{equation}
$u = r/R_d$,} calculated using the fitting formula provided by \citep{mo:1998a},
\begin{equation}
F_R\approx \left( \frac{j_d \lambda}{0.1 m_d}\right)^q (1-3m_d +5.2 m_d^2) g(c),
\end{equation}
where $q={-0.06 +2.71 m_d +0.0047(m_d/j_d\lambda)}$, and $g(c) = (1-0.019c+0.0002c^2+0.52/c)$.
Using the virial theorem, the total energy of the halo is:
\begin{equation}
E = - K = - \frac{M_{vir} v_c^2}{2} F_E,
\label{eq:E}
\end{equation}
where the factor $F_E$ encloses deviations from the energy of an isothermal sphere ($E = -\frac{1}{2}M_{vir}v_c^2$ assuming all particles to be in circular orbits), and depends on the exact shape of the halo density profile; for the NFW profile one finds
\begin{equation}
F_E = \frac{c}{2F^2(c)} \left[ 1 - \frac{1}{\left(1+c\right)^2} - \frac{2 \ln(1+c)}{1+c}\right].
\label{eq:F_E}
\end{equation}

Combining eqs. \ref{eq:lambda}, \ref{eq:md,jd}, \ref{eq:J_d} and \ref{eq:E},  we finally find $R_d$ in terms of $m_d$ and $j_d$:
\begin{equation}
R_d = \frac{\lambda}{\sqrt{2}} \left( \frac{j_d}{m_d} \right) F_R^{-1} F_E^{-1/2} r_{vir}.
\label{eq:R_d}
\end{equation}

{ We assumed the disk to have a constant vertical density $\rho_d (R, z) = \Sigma(R) H^{-1}$, for $\lvert z \rvert \leq H$, and $\rho_g$ (Eq. \ref{eq:makino}), otherwise.
The scale height $H$ of the disk
 can be easily calculated if hydrostatic equilibrium is assumed\footnote{ This estimate neglects that a rigorous calculation of $H$ should account for the fact that the constant temperature and density vertical profile assumed for the disk do not allow hydrostatic equilibrium to occur.}:
\begin{equation}
\frac{dP}{dz} = - \rho g_z
\label{eq:idro}
\end{equation}
{ Then, if we assume the gravity is dominated by the halo, the gravitational accerelation at a distance $\sqrt{R^2 + z^2}$ from the disk center can be approximated by:}
\begin{equation}
g = \frac{G M(R) }{R^2 + z^2},
\end{equation}
with
\begin{equation}
g_z = \frac{G M(R) }{R^2 + z^2} \sin \theta = \frac{G M(R) }{R^2 + z^2} \frac{z}{\sqrt{R^2+z^2}}
\end{equation}
where $\theta$ is the angle between $\bf {g}$ and $\bf{r}$.
At the disk edge, $R = R_d > z$:
\begin{equation}
g_z \approx \frac{G M(R) z}{R_d^3},
\end{equation}
and Eq. \eqref{eq:idro}:
\begin{equation}
dP = - \rho \frac{G M(R_d) z}{R_d^3} dz.
\end{equation}
{ For $R = R_d$, the $z$-dependence of $\rho$ can be neglected and the integration between 0 and $H$ gives:
\begin{equation}
P \approx \rho c_s^2 \approx \rho \frac{G M(R_d)}{R_d^3} \frac{H^2}{2},
\end{equation}
so that
\begin{equation}
H \approx \sqrt{2\frac{c_{s,d}^2 R_d^3}{GM_d}} \approx \frac{R_d}{6}.
\label{eq:H}
\end{equation}
}

\subsection{Black Hole accretion}
\label{ssec:accretion}
The accretion theory of a point mass (i.e. a BH) in motion through a uniform medium has been developed by
\citealt{bondi:1952, bondi:hoyle:1944, hoyle:lyttleton:1939}. The accretion rate $\dot{M}_{\bullet}$ depends on 
both the BH-gas relative velocity\footnote{ In our approximation the gas in the galaxy is static and $v_\bullet$ coincides with the BH velocity. Nevertheless, even if $v_\bullet$ should account for the rotational velocity of the gas disk, the wide range of simulated BH velocities (sec. \ref{sec:method}) covers the discrepancy from the BH-gas relative velocity.}, $v_\bullet$, and the gas density, $\rho$. In practice, radiation emitted in the accretion process limits the accretion itself by exerting a radiation pressure onto the infalling gas. Hence, the maximum accretion rate achievable, $\dot{M}_{Edd}$, is the one that produces the Eddington luminosity 
$L_{Edd} = 1.5 \times 10^{38} (M_\bullet/M_\odot) \, \rm erg \, s^{-1}$.

To account for feedback-limited accretion we build upon the results by 
\cite{park_ricotti:2013} (hereafter PR; see also \citealp{park_ricotti:2011, park_ricotti:2012})
obtained assuming spherically symmetric accretion and a radiative efficiency $\eta =  L/\dot M c^2 =0.1$.
In their simulations, PR find that the accretion rate is strongly dependent on $v_\bullet$, depending on which three different regimes have been identified:
{
\begin{equation}
\dot{M}_{\bullet} = \begin{cases}
0.01 T_4^{5/2} n_5^{1/2} \dot{M}_B & 
v_\bullet<c_s \\
0.7 \frac{G^2 M_\bullet^2}{c^3_{s,in}} \rho ({v_\bullet}/2 c_{s,in})^2 &
c_s<v_\bullet<2 c_{s,in} \\
\frac{G^2 M_\bullet^2}{c^3_{s,in}} \rho (1+v_\bullet^2/c_{s,in}^2)^{-3/2} &
v_\bullet \gg 2 c_{s,in},
\end{cases}
\label{eq:parkricotti}
\end{equation}
}where $\dot{M}_B = \pi e^{3/2} \rho G^2 M_\bullet^2 c_s^{-3}$ is the Bondi rate, $n = \rho/\mu m_p$,
and $c_{s_{in}} = (\gamma k_B T_{in}/\mu_{in} m_p)^{1/2}$ is the sound speed in the cometary-shaped HII 
region around the BH. We assume $\mu_{in} = 0.6$ (fully ionized primordial gas) and $T_{in} = 4\times 10^4 \rm K$, corresponding to a spectral index $\alpha = 2$ if a single power law $\nu^{-\alpha}$ for the black hole emission spectrum is assumed \citep{park_ricotti:2011}. As can be appreciated from Fig. \ref{fig:parkricotti}, the accretion rate {decreases by several orders of magnitudes as $v_\bullet > c_s$ (i.e. the motion is supersonic with respect to the ambient medium), when PR predicts the formation of a dense gas shell behind the ionization front. The accretion rate has another discontinuity for $v_\bullet = 2 c_{s_{in}}$, when the ionized layer becomes rarefied.}

\begin{figure}
\centering
\includegraphics[width=0.45\textwidth]{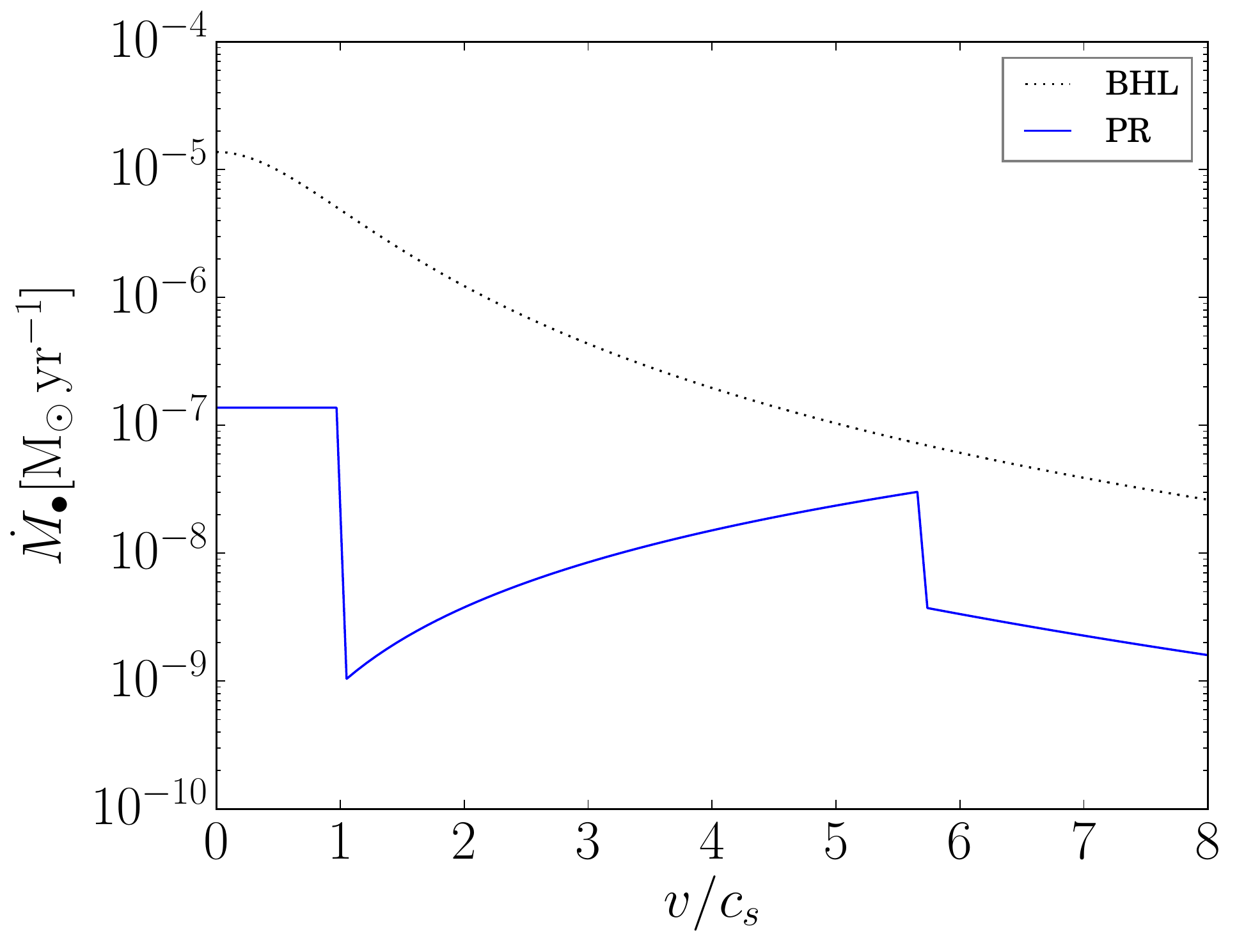}
\caption{ BH accretion rate $\dot{M}_\bullet$ from eq. \ref{eq:parkricotti} (PR) compared to the Bondi-Hoyle-Lyttleton model (BHL) eq. \ref{eq:BHL}. An ambient gas density $n = 10^{5} \, \rm cm^{-3}$, temperature $T = T_{vir} = 10^4 \; \rm K$ and $M_\bullet = 100 M_\odot$ are assumed.}
\label{fig:parkricotti}
\end{figure}

The previous formulae implicitly assume that the accretion flow on the black hole is spherically symmetric. However, under
some conditions, it is possible that in the vicinity of the black hole the gas is funneled into an accretion disk. 
The most important condition leading to an accretion disk formation is that the captured material specific angular momentum is larger than the one of the innermost stable orbit $J_{ISCO} = \sqrt{3} c r_g$ for a Schwarzschild black hole ($r_g = 2GM_\bullet/c^2$ is the gravitational radius).

Although disk formation around a moving black hole is still controversial (see for example \cite{chisholm:2003} and \cite{davies:1980} for arguments in favor of disk-like or spherical geometry, respectively) we adopt a spherical accretion model
on the basis of the arguments given in \cite{beskin:karpov:2005}. These authors conclude that, for
nearly any $v_\bullet$, neither ISM turbulent motions nor density fluctuations 
can prevent the onset of a spherical accretion regime.
As an example of stellar-BHs disk-like accretion with the PR accretion model, we nevertheless refer to \cite{wheeler:2011}.

\subsection{Emitted spectrum}
\label{ssec:spectrum}
As the black hole accretes and grows, part of the flow kinetic energy is transformed into radiation. We start by noting that, in a spherical-accreting plasma bremsstrahlung emission has been shown to occur with a radiative efficiency $\eta \simeq 10^{-4}$ \citep{shapiro:1973}, that is very low compared with the one 
achievable in a disk-like accretion\footnote{About $6\%$ for a Schwarzschild (i.e. non-rotating) black hole to
more than $30\%$ for a maximally rotating one \citep{thorne:1974}.}.
Moreover, \cite{shvartsman:1971} shows that for $M_\bullet < 10^4 M_\odot$ the gas density in the accretion flow is so low that the cooling time is longer than the free fall time.

For low accretion rates the dominant radiative process is synchrotron emission.
Indeed, there is growing evidence that strong magnetic fields are present in high-$z$ galaxies \citep{bernet:2008, robishaw:2008, murphy:2009}; they are thought to be generated through turbulent amplification of weak, relic magnetic seed fields \citep{gnedin:2001,sur:2010,federrath:2011,banerjee:2012}, generated during the very early stages of cosmic evolution \citep{turner:1988, durrer:2013}. 
The evaluation of the emitted spectrum is particularly complex, involving the calculation of the particle energy distribution in the inflowing plasma.

The particle energy distribution is determined by the superposition of three contributions:
adiabatic heating, synchrotron cooling and non-thermal heating, the latter being a consequence of the 
\cite{shvartsman:1971} equipartition theorem of thermal, gravitational and magnetic energy density.
For this equipartition to be valid, there must be a dissipation mechanism of the magnetic field, which therefore is no longer frozen-in: the conductivity decreases, relative motions between the magnetic field and the plasma develop, and currents begin to flow and heat the gas. Such process dissipates about $25\%$ of the rest energy of the infalling material \citep{beskin:karpov:2005} as a result of the formation of turbulent current sheets that cause fast magnetic field lines reconnection and energy release\footnote{Magnetic reconnection is also invoked to explain solar flares.}. 
About $10\%$ of the energy released  in reconnection events goes into excitation of plasma oscillations,
whereas the remaining $90\%$ accelerates particles up to a maximum Lorentz factor $\gamma_{max} \simeq 10^5$.

In conclusion, the total particle energy distribution\footnote{The distribution is not normalized and only the shape has physical meaning.}  can be expressed as the sum of a thermal and a non-thermal component:
\begin{equation}
f(R,\gamma) = f_t(R, \gamma) + \zeta f_{nt}(R, \gamma)  \propto \frac{d^2N}{dR d\gamma},
\label{eq:distrib}
\end{equation}
where $R = r/r_g$, and $\zeta$ expresses the ratio between the non-thermal and thermal electrons densities:
\begin{equation}
\frac{n_{nt}(R)}{n_t(R)} = \zeta \frac{f_{nt}(R)}{f_t(R)}
\end{equation}
and $f(R) = \int f(R, \gamma) d\gamma$.
For the derivation of an analytical expression of $f_t$ and $f_{nt}$ we refer the reader to \cite{beskin:karpov:2005}.

Now consider a particle free-falling in the black hole. Knowing the (isotropic) emissivity of the particle in its own rest frame, $j_{\nu '}$, and accounting for relativistic time contraction, gravitational redshift, Doppler effect and the fraction of the emission disappearing in the event horizon, we can compute the single-electron emission spectrum seen from an observer at infinity \citep{shapiro:1973}. This is:
\begin{equation}
L_\nu^{e^-} = 2 \pi \int_{-1}^{\cos{\theta^*}} j_{\nu '} \frac{1 - \beta^2}{\left( 1-\beta \cos{\theta}\right)^2} d\cos\theta \rm \; \; erg \; s^{-1} \, Hz^{-1},
\label{eq:lum}	
\end{equation}
where
\begin{equation}
|\cos \theta^* | = \sqrt{1-\frac{27}{4R^2}\left( 1- \frac{1}{R}\right)}
\end{equation}
is the event horizon angular size for a free-falling emitter \citep{zeld:1971},
\begin{equation}
\beta = \frac{dr}{dt} \frac{1}{1-r_g/r} = \frac{v/c}{\left(v^2/c^2 + 1 - r_g/r\right)^{1/2}}
\end{equation}
is the free-falling velocity of the particle in the distant observer rest frame, and the frequency shift is:
\begin{equation}
\nu ' = \nu   \frac{1- (v/c) \cos\theta}{\sqrt{1-(v^2/c^2)(1-1/R)}}.
\end{equation}
The synchrotron emissivity\footnote{We implicitly assume the particle velocity to be orthogonal to the magnetic field: in fact adiabatic compression increases only the perpendicular component of the particle momentum.} is \citep{rybicki:1979}:
\begin{gather}
j_\nu =   \frac{\sqrt{3}e^3B}{4\pi m_e c^2} F\left(\frac{\nu}{\nu_c}\right) \\
F(x) = x\int^\infty_x K_{5/3}(\xi)d\xi \\
\nu_c = \frac{3 \gamma^2 e B }{4 \pi m_e c},
\end{gather}
where $K_{5/3}$ is the modified Bessel function, and 
\begin{equation}
B = 8\times 10^4 \left( \frac{10^5\dot{M}_\bullet}{\dot{M}_{Edd}}\right)^{1/2} \left( \frac{M_\bullet}{10 M_\odot}\right)^{-1/2} R^{-5/4} \rm G
\end{equation}
for the Shvartsman equipartition theorem.
{ $\dot{M}_{Edd} = L_{Edd}/c^2 = 2.7 \times 10^{-7} (M_\bullet/100 M_\odot) \, M_\odot \rm \, yr^{-1}$ is the Eddington rate.}

With the electron energy distribution in Eq. \eqref{eq:distrib},  the accretion flow emission spectrum is then:

\begin{equation}
L_\nu = 2\sqrt{3} \pi \frac{r_g^2}{\sigma_T} \left( \frac{\dot{M}_\bullet}{\dot{M}_{Edd}} \right) \int_1^\infty dR \; \int_1^\infty d\gamma \; f(R,\gamma) \; L_\nu^{e^-} ,
\label{eq:sync_spec}
\end{equation}
where $\sigma_T$ is the Thomson cross section and the dependence on the black hole mass is in the gravitational radius $r_g$.
This model predicts the synchrotron emission of an accreting BH up to accretion rates $\dot{M}_\bullet \leq 10^{-4} \dot{M}_{Edd}$ \citep{beskin:karpov:2005}. In the few cases in which we derive larger accretion rates, we conservatively renormalize the spectrum so that the radiative efficiency $\eta< 20$\%.
The spectral energy distribution of a $100 M_\odot$ BH for different accretion rates $\dot{M}_\bullet$ is shown in Figure \ref{fig:sed}.

\begin{figure}
\centering
\includegraphics[width=0.45\textwidth]{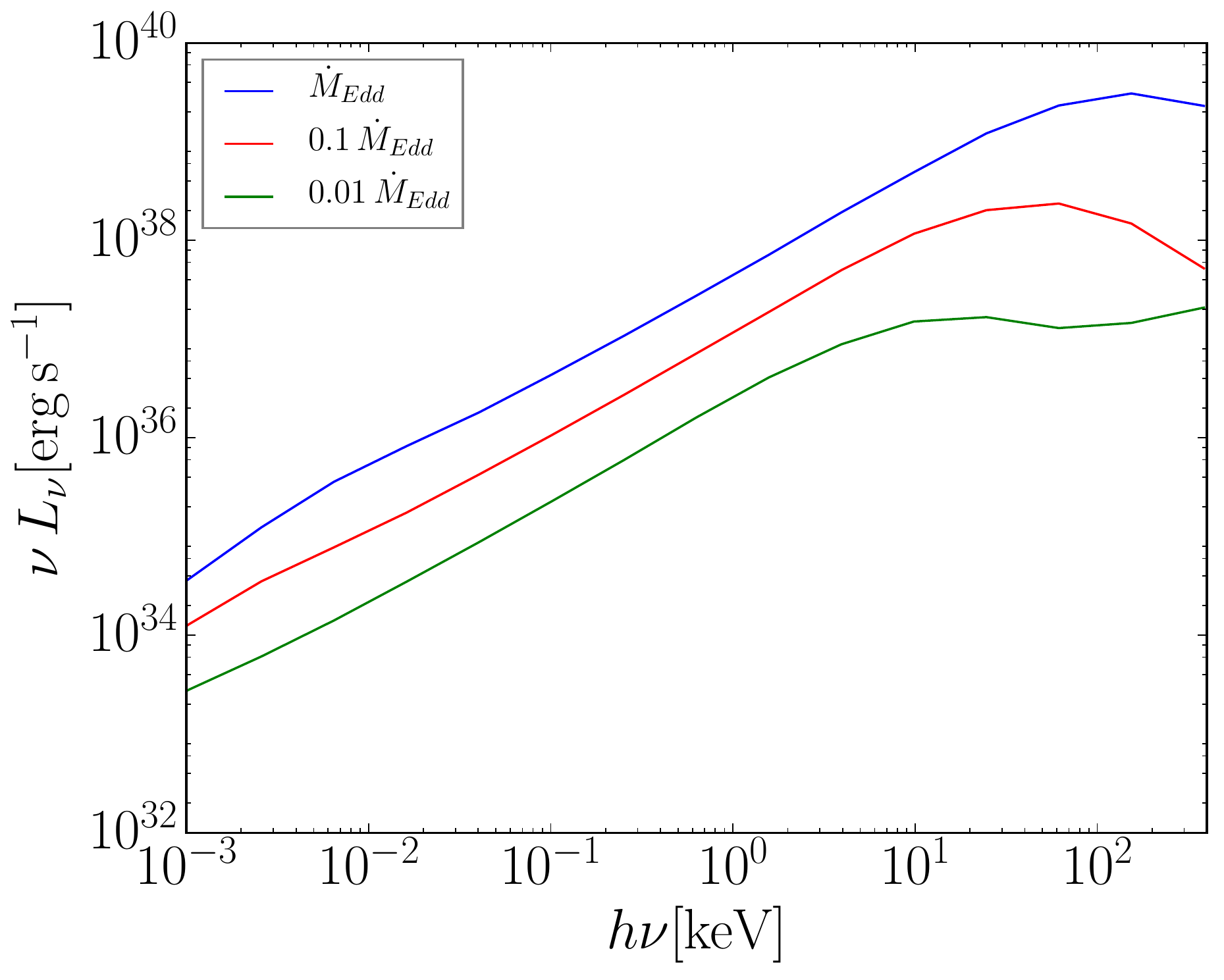}
\caption{Synchrotron spectrum (eq. \ref{eq:sync_spec}) emitted by gas infalling onto a $100 M_\odot$ BH for different values of the accretion rate $\dot{M}_\bullet$. }
\label{fig:sed}
\end{figure}

\subsection{Black hole dynamics}
\label{sec:method}
As discussed in sec. \ref{ssec:accretion}, the accreted mass, $\Delta M$, depends on the gas density, $\rho$, and the BH velocity, $v_\bullet$, at each point along its orbit, in turn determined by the galaxy gravitational potential produced by the assumed matter distribution (Sec. \ref{ssec:host_gal}). We start by writing the equation of motion which yields the BH acceleration, 
\begin{equation}
{\ddot{\bf r}} = - \nabla \phi(r) - \frac{\dot{M}_\bullet}{M_\bullet} {\bf v}_\bullet,
\label{eq:acceleration}
\end{equation}
where the gravitational potential, $\phi$,  is given by the Poisson equation:
$ \nabla^2 \phi = 4 \pi G \rho$. 
The gravitational potential contains contributions from two components: (a) the spherically symmetric distribution of the dark matter halo (potential $\phi_h$), and (b) the baryonic disk ($\phi_d$). We neglect the subdominant contribution from the fraction of uncollapsed baryons residing in the halo.   

The halo-related potential is  simply
\begin{equation}
\phi_h(r) = - \frac{G M(r)}{r}.
\label{eq:phi}
\end{equation}
To compute the potential of the second component we adopt a thin-disk approximation, yielding 
\begin{equation}
\phi(r,z)_{d} = -2 \pi G \int^\infty_0 J_0(kr) \hat{\Sigma}(k) e^{-k|z|} dk,
\label{eq:thin_disk}
\end{equation}
where $J_0 (x)$ is the zero-th order Bessel function, and $\hat{\Sigma} (k) = \int^\infty_0 J_0(kr) \Sigma(r) r dr $
is the Hankel (or Fourier-Bessel) transform of $\Sigma$ \citep{toomre:1963}. Finally, by combining eqs. \ref{eq:thin_disk} and \ref{eq:surf_density} we get
\begin{equation}
\phi(r,z)_{d} = -2 \pi G \Sigma_0 R_d^2 \int^\infty_0 \frac{J_0(kr) e^{-k|z|}}{[1+(kR_d)^2]^{3/2}} dk.
\label{eq:pot_disk}
\end{equation}
The plot of the potentials for the two components, along with the total one, $\phi = \phi_h + \phi_d$ in the disk plane as a function of galactocentric radius, $R$, is shown in Fig. \ref{fig:pot}.

\begin{figure}
                \centering
                \includegraphics[width=0.45\textwidth]{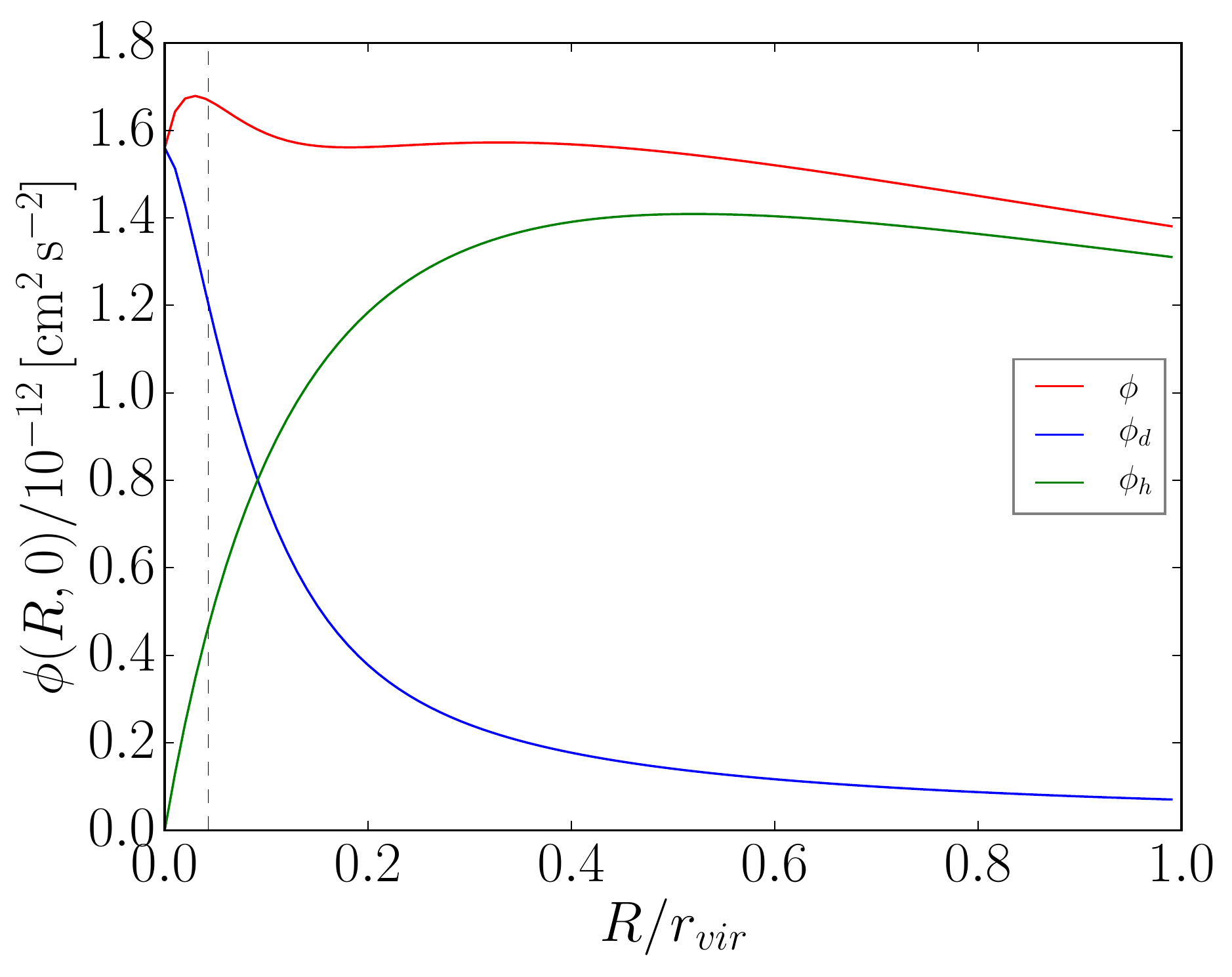}
                \caption{Total gravitational potential $\phi$ (red) as a function of galactocentric radius in the disk plane. Also shown are the disk (blue) and halo (green) contributions.  The vertical dashed line marks the disk scale length $R_d$.}
                \label{fig:pot}
\end{figure}
\begin{figure}
                \centering
                \includegraphics[width=0.45\textwidth]{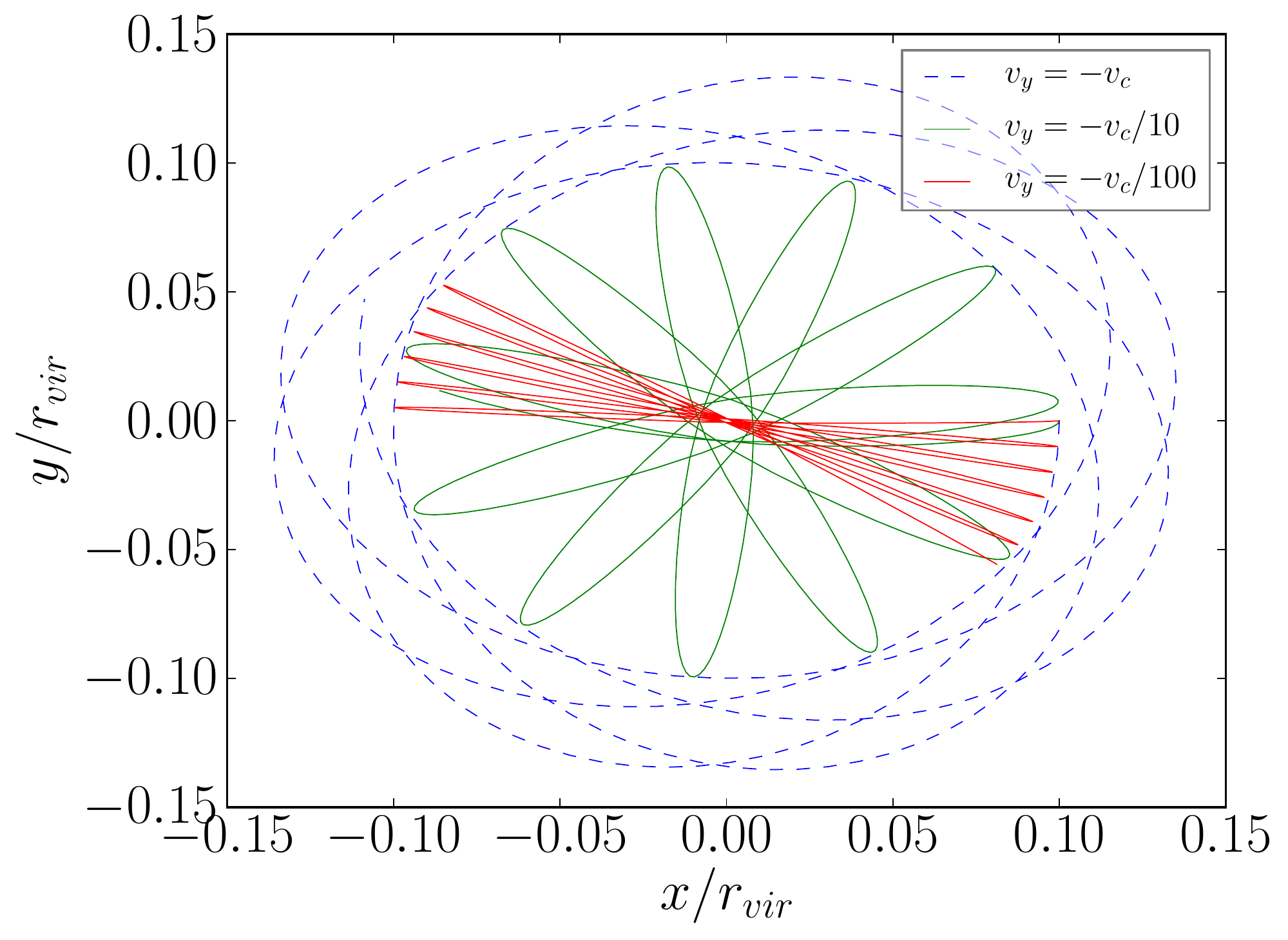}
                \caption{Example of orbits in a spherical galaxy hosted by a $T_{vir} = 10^4 \rm K$ DM halo, starting from different initial conditions.}
                \label{fig:orbits}
\end{figure}

The trajectory of a particle in a given potential well is completely determined by three parameters (reduced to two for spherically symmetric problems). Given the initial conditions (IC), we solve eq. \ref{eq:acceleration} using a leapfrog integration
over a total integration time\footnote{We recall that the age of the universe is $5\times10^8 \rm yr$ and $8\times10^8 \rm yr$ at $z=10$ and $z=7$, respectively.}, $t_\mathrm{max}=300 \, \rm Myr$ with a typical time step of $dt = 10^3 \rm yr$.

For the spherical case, we consider without loss of generality all orbits to lie on a plane $(x,y)$ perpendicular to the rotation axis $z$. We initialize the BH position as $x =r_0$, $y=0$, with initial velocity components $v_x = 0$, $v_y = v_0$.
The values of $r_0$ and $v_0$ are bound to be $0 < r_0 < r_{vir}$ and $0 < v_0 < v_c$, respectively. 
The second bound implies that the initial velocity is smaller than the escape velocity in Eq. \eqref{eq:v_e}, and the orbital period $T_r \simlt t_\mathrm{max}$.
We verified that, with these IC, the BH never enters in the accretion regime $v_\bullet > 2 c_{s_{in}}$ (see eq.  \ref{eq:parkricotti}).
Fig. \ref{fig:orbits} shows an example of BH orbits in a spherical galaxy hosted in a $T_{vir} = 10^4 \rm K$ halo, starting from $r_0 = 0.1 \,  r_{vir}$
and $v_0 = [1, \; 0.1, \; 0.01] v_c$. 

When a disk is present in addition to the halo, spherical symmetry does not hold. Then we need to specify three, rather than two, orbital parameters. These are: ${\bf r}_0 = (x_0, 0, x_0 \tan \alpha), {\bf v_0} = (0, v_0, 0)$. Unless differently stated, we take the angle $\alpha = 0$. For simplicity, and to allow a direct comparison between the spherical and the disk+halo cases, no radial component of the initial velocity is considered in this study. 

\section{Results}
\label{sec:results}
We applied the method described in Sec. \ref{sec:method} to evaluate the mass accreted by a $100 M_\odot$ black hole as a function of time along its orbit. We are in particular interested in assessing how the rate depends on the IC of the motion and whether significant growth can occur.  We have investigated halo virial temperatures spanning the extended range $5000 < T_{vir} < 10^5 \rm K$. However, we found very weak dependence of the results on $T_{vir}$; for sake of brevity and in view of the conclusions drawn, we then limit the following discussion to the case $T_{vir} = 10^4 \rm K $ only.

\subsection{SPHERICAL GALAXIES}
\label{ssec:spherical_galaxy}
Our investigation allows a statistical description of the mass accreted by the 100 $M_\odot$ BH during the integration time $t_\mathrm{max}=300$ Myr as a function of the orbital ICs. We recall that this time span implies that for a BH observed at $z=7$ the accretion phase has initiated at $z= 10$. For the spherical galaxy case analyzed in this Section, the results are summarized in the left panel of Fig. \ref{fig:Sas}.

\begin{figure*}
   \centering
   \includegraphics[width=0.495\textwidth]{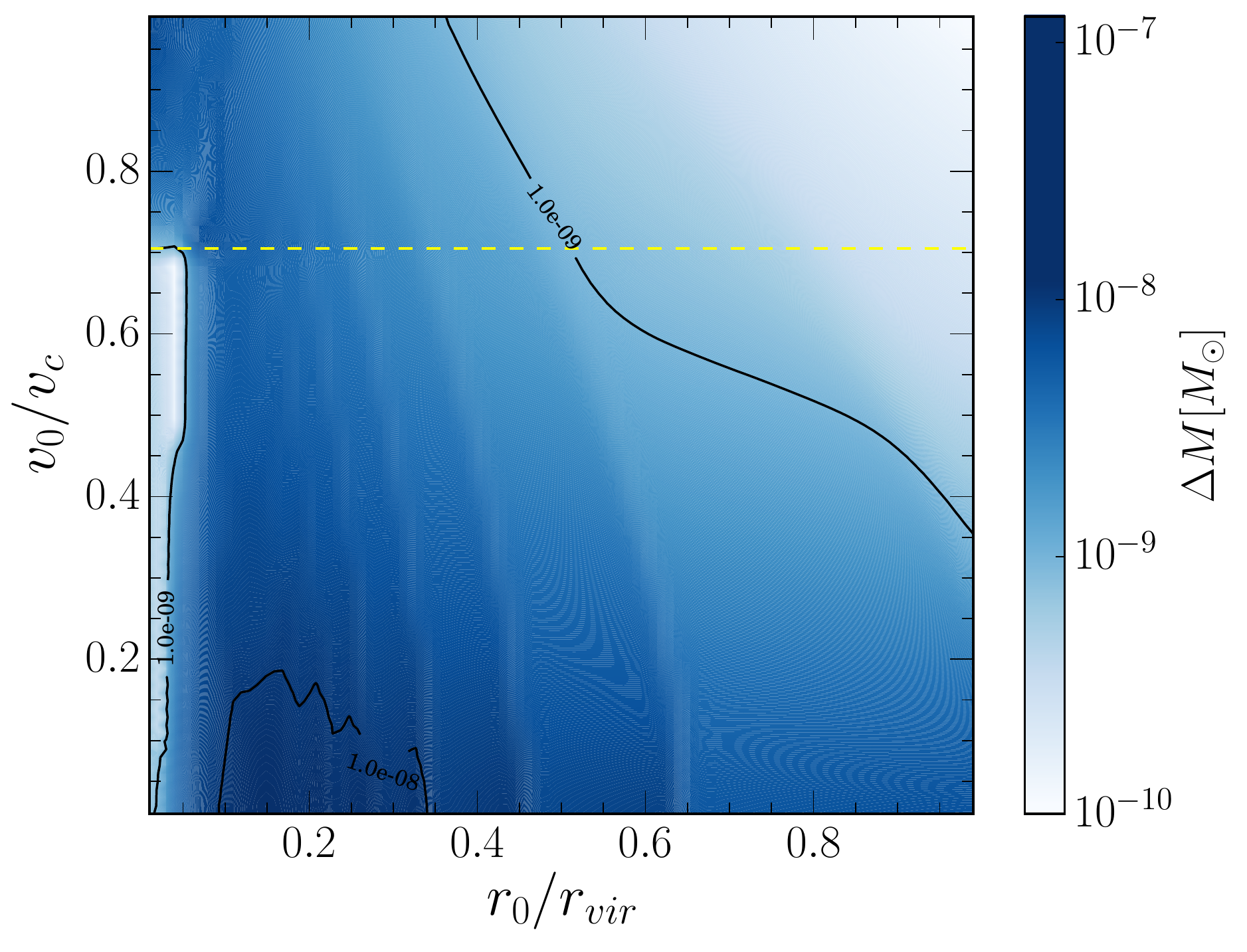}
   \includegraphics[width=0.49\textwidth]{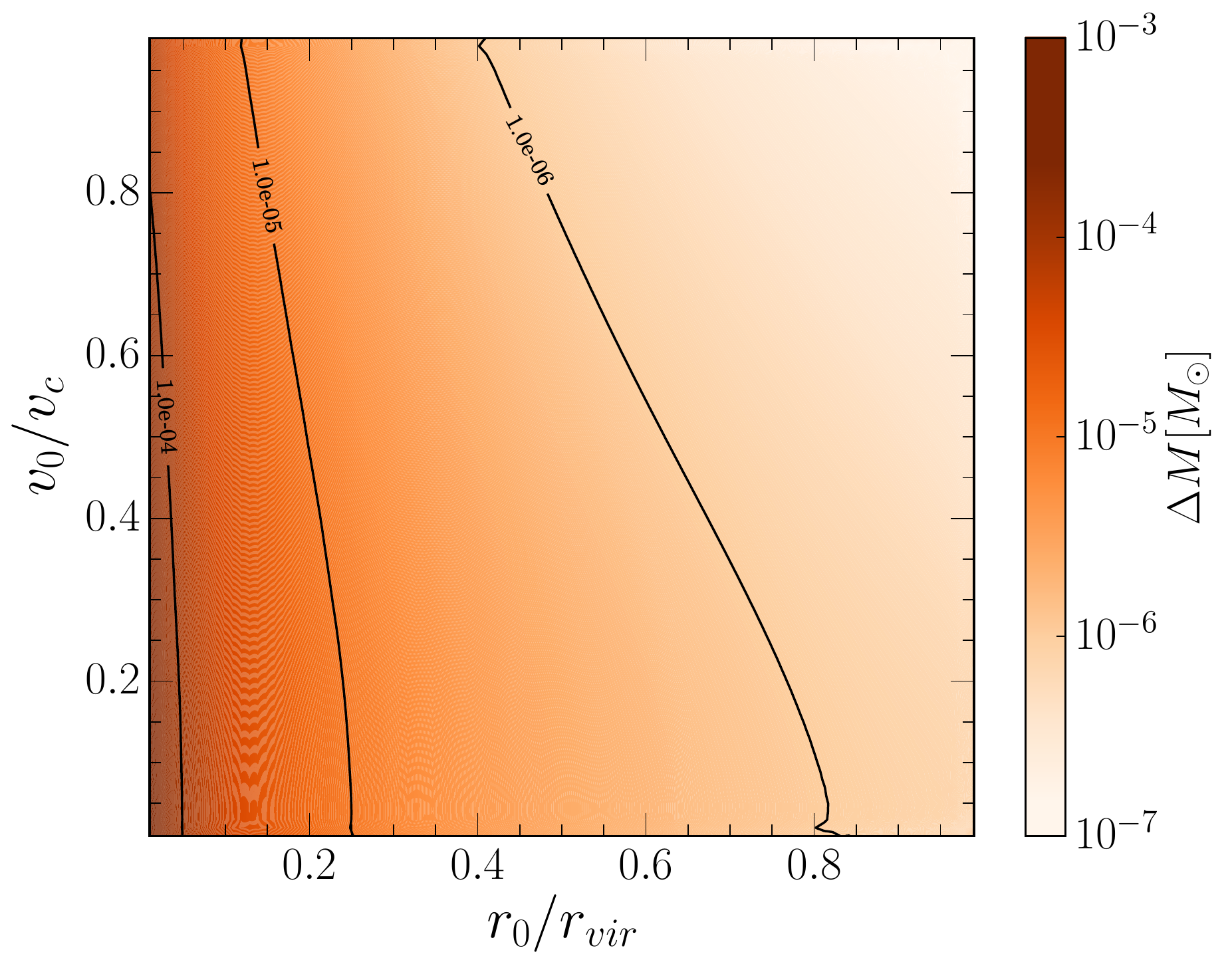}
                \caption{\emph{Left panel:} mass accreted in 300 Myr by a $100 M_\odot$ BH (spherical galaxy model) as a function of the orbital ICs $(r_0, v_0)$. The yellow dashed line corresponds to the sonic point at which $v_0 = c_s$; below (above) the line the BH is subsonic (supersonic) corresponding to efficient (inefficient) accretion as shown in Fig. \ref{fig:parkricotti}. Maximal accretion ($\Delta M \simlt 10^{-7} M_\odot$) is attained when the BH traverses at subsonic speed the central regions of the galaxy; this mainly occurs for close subsonic ICs. { Black contour lines join points of equal $\Delta M$ ($10^{-9}$ and $10^{-8} M_\odot$ in this case).} \emph{Right panel:} same as left panel but with radiative feedback turned off. Accretion occurs at the standard Bondi-Hoyle-Lyttleton rate. $\Delta M \simlt 10^{-3} M_\odot$ is much higher than in the model included feedback, but still negligible with respect to the initial BH mass.}
\label{fig:Sas}
\end{figure*}

The accretion rate is maximal when the BH moves at subsonic speed into the densest central regions of the galaxy. 
However, even under these favorable conditions we find that the mass gained during the evolution is very limited, 
amounting to only $10^{-7} M_\odot$, certainly insufficient to build a SMBH by $z \approx 7$. In these cases, the mean accretion rate is at most $\langle\dot{M}_\bullet\rangle = \Delta M / t_{max} < 10^{-8} \dot{M}_{Edd}$. The rest of the parameter space, 
with ICs corresponding to central subsonic orbits or distant supersonic ones, virtually does not allow any growth 
($\Delta M \approx 10^{-10} M_\odot$).

It is interesting to check whether such BH inability to accrete matter descends from feedback. The answer is encoded in the right panel of Fig. \ref{fig:Sas}, in which we have repeated the same experiment shutting off radiative feedback. A standard Bondi-Hoyle-Lyttleton accretion rate \citep{bondi:1952, bondi:hoyle:1944, hoyle:lyttleton:1939}, 
\begin{equation}
\dot{M}_{BHL} =  \pi e^{3/2} \frac{G^2 M^2_\bullet }{\left(c_s^2 + v^2_\bullet \right)^{3/2}}\rho,
\label{eq:BHL}
\end{equation}
has been assumed in this case. Although indeed feedback causes a quenching of the accretion rate by 3-4 orders of magnitude, even neglecting its effects does not allow a sensible growth of the BH ($\Delta M \approx 10^{-3} M_\odot$). Thus, we can conclude that in galaxies not developing a central disk condensation, BH growth is essentially impossible. Hence, we next examine the disk galaxy case.

\subsection{DISK GALAXIES}
\label{ssec:disk_galaxy}

In principle, the high gas densities (up to $n \approx 10^3 \cc$) found in the disk favor high accretion rates. However, the limiting factor to such efficient accretion is the residence time of the BH in the disk. The maximum accretion rate is expected for orbits in which the residence time is 100\%; these are the orbits lying in the disk plane ($\alpha = 0$). The results of this particular case are shown in the left panel of Fig. \ref{fig:as_disk}. In this case, much larger accretion rates with respect to the spherical galaxy case can take place, leading to a small maximal mass growth $\Delta M \approx 10^{-2} M_\odot$,
so that the mean accretion rate is at most $\langle\dot{M}_\bullet\rangle = \Delta M / t_{max} < 10^{-3} \dot{M}_{Edd}$.
{ The low-$\Delta M$ region in Fig. \ref{fig:Sas} is the combination of two closely related factors: (i) the BH accretion is efficient (subsonic regime) for about 3\% of the integration time, (ii) its trajectory crosses a density larger than $ 10^{-26} \, \rm g cm^{-3}$ for 3\% of the time only.
This region does not appear in the disk case (Fig.\ref{fig:as_disk}) where, even if the accretion regime is inefficient as well, the BH travels in denser ($> 10^{-26} \rm \, g cm^{-3}$) regions for at least 20\% of the integration time.
}

If the orbit is instead inclined at an angle $\alpha$ with respect to the disk plane the situation changes only slightly, as shown in Fig. \ref{fig:ang_dep}. We note that the BH trajectory is vertically contained in the baryonic disk if the inclination angle is $\alpha_d = \tan^{-1}(H/R_d) \approx 10^o $; thus, smaller inclination angles give the same result as the $\alpha=0$ one. 
Depending on the initial position and velocity, small ($\simlt 3$) variations are predicted (Fig. \ref{fig:ang_dep}) as a function of $\alpha$. 
In particular, orbits starting near the center with low velocity show an increase in $\Delta M$ for large inclinations, since the BH spends a larger fraction of the integration time in an efficient accretion regime (as can be deduced from Eqs. \eqref{eq:phi} and \eqref{eq:pot_disk}, the BH experiences a stronger acceleration for large inclinations).

As already noted, velocity seems to play the most important role, with subsonic orbits leading to somewhat larger accretion rates and mass growth, yet still largely insufficient to support the idea that SMBH can grow out of stellar mass BHs orbiting in primordial galaxies.  

Finally, we discuss the case in which a standard Bondi-Hoyle-Lyttleton accretion rate (eq.\ref{eq:BHL}) is assumed (Fig. \ref{fig:as_disk}, right panel). As in the spherical case, radiative feedback quenches accretion by 3-4 orders of magnitude and, even if an increment $\Delta M \approx 35 M_\odot$ can be achieved under the most favorable ICs, this is largely insufficient to grow a SMBH from a $100 M_\odot$ one.


\begin{figure*}
   \centering
   \includegraphics[width=0.495\textwidth]{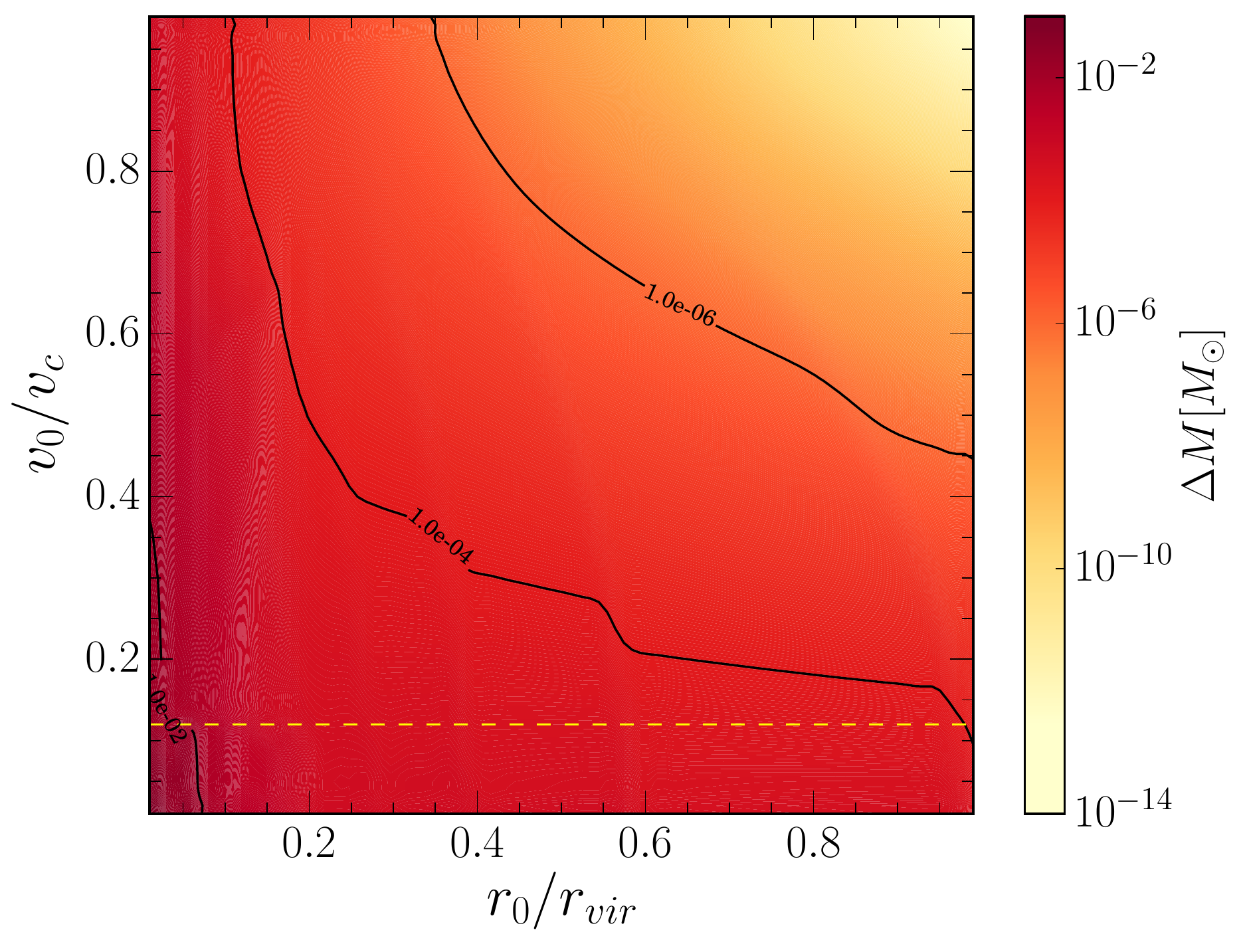}
   \includegraphics[width=0.49\textwidth]{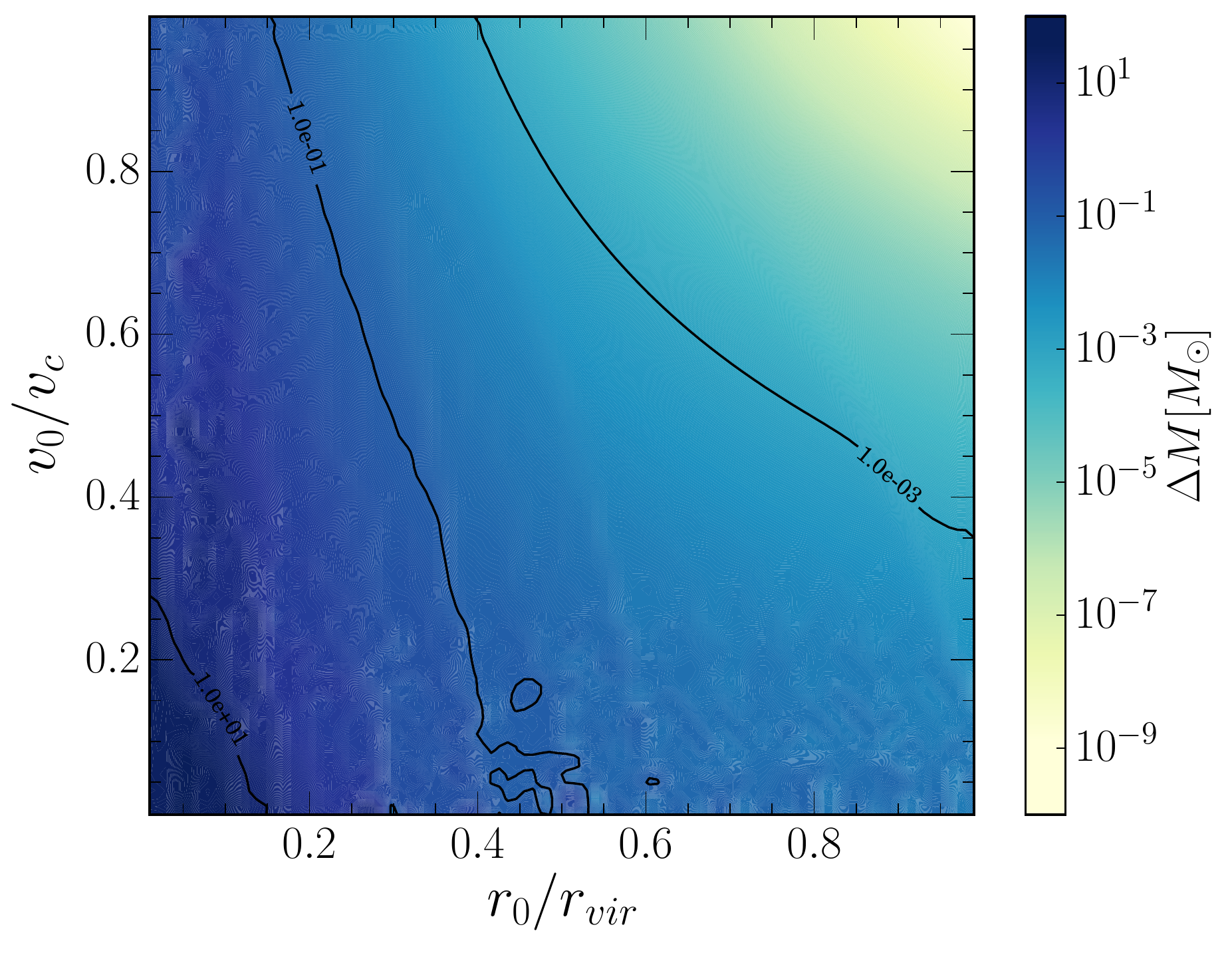}
                \caption{Same as Fig. \ref{fig:Sas}  but for the disk galaxy case.}
 \label{fig:as_disk}
\end{figure*}

\begin{figure}
                \centering
                \includegraphics[width=0.49\textwidth]{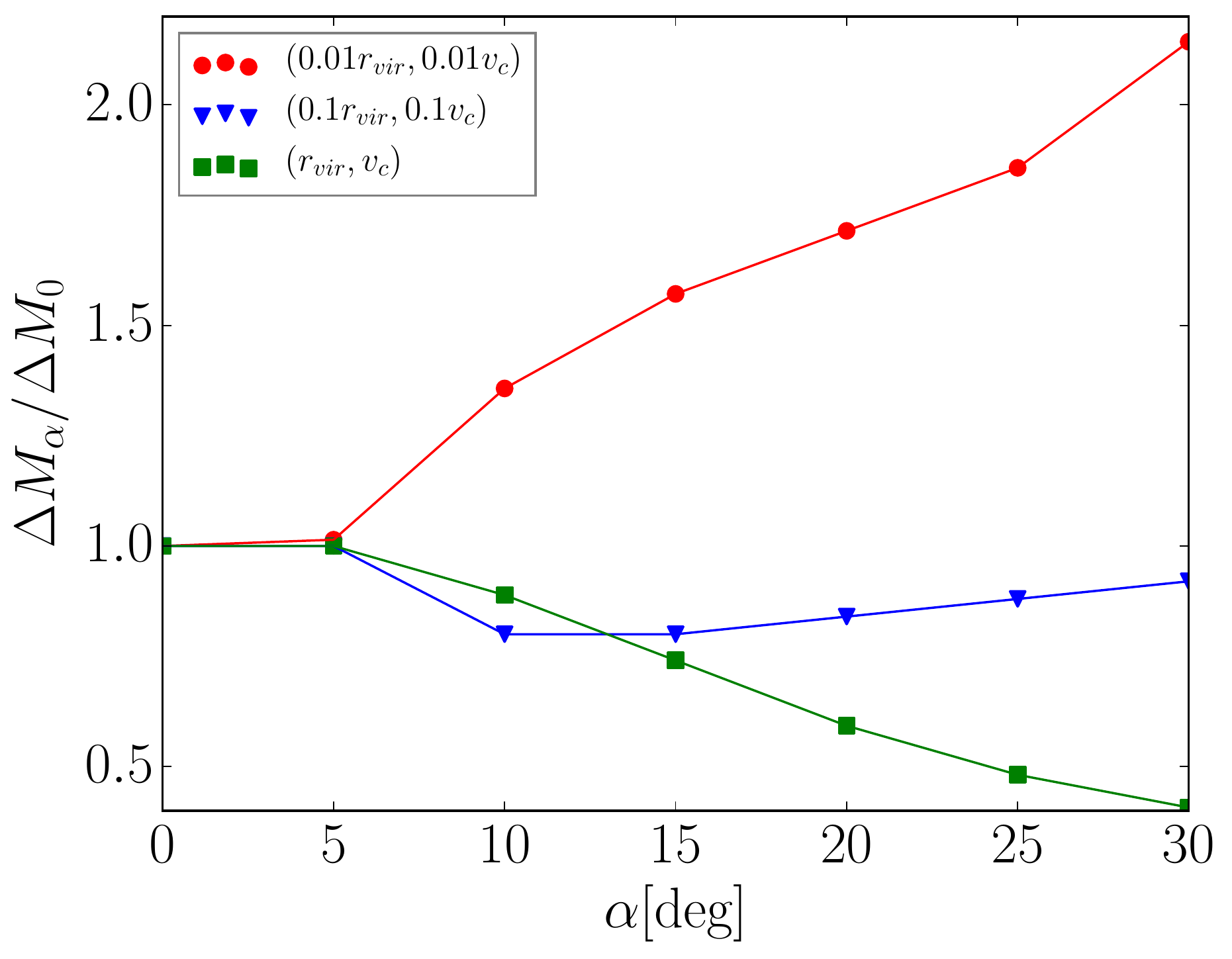}
		\caption{Relative mass increment for orbits at different inclination angle $\alpha$ with the disk plane with respect to the orbit at $\alpha=0$ for different initial conditions, $(r_0, v_0)$ as shown in the inset.}
                \label{fig:ang_dep}
\end{figure}

\section{Emission}
\label{sec:emission}
Although we have seen in the previous Sections that, under the conditions explored here, BH growth is extremely slow, it is nevertheless important to predict what level of radiative emission is produced during the evolution. 
X-ray photons produced by early galaxies ionize neutral atoms in the IGM, ejecting high-energy photoelectrons that ionize or collisionally excite other atoms whose energy is eventually thermalized by scattering with other electrons. Such process deposits 1/3 of the emitted energy as heat \citep{shull:1985, furlanetto:2010, valdes:2007}.
This preheating of the neutral medium 
can leave well defined signatures on the 21 cm line power spectrum ( e.g. \citealt{mesinger:2013, evoli:2014}).

In this Section, we want to clarify whether X-ray emission of stellar BHs in primordial galaxies can produce detectable signature on the 21-cm signal from the high-$z$ Universe.
We focus on the disk galaxy case since accretion rates are larger, and consequently emission is more conspicuous. 

The number $N_\bullet$ of stellar mass BHs in the galaxy can be estimated assuming that stars form according to a 
Salpeter initial mass function (IMF) $\phi(m) \propto m^{-2.35}$ between $(1-500) M_\odot$. 
Stars with $m > 8 M_\odot$ explode as supernovae with a frequency $\nu_{SN} \approx 1/100 M_\odot$;
among them, those more massive than $m> 25 M_\odot$ leave a black hole remnant.
According to \cite{heger:2002}, stars with mass between $25$ and $140 M_\odot$ leave BHs between $2$ and $100 M_\odot$, while stars heavier than $260 M_\odot$ leave BHs up to $500 M_\odot$.
Between $140$ and $260 M_\odot$ theoretical forecasts predict stars  to explode as pair-instability supernovae, leaving no remnants.

With these assumptions, in a DM halo of virial mass  $M_{vir} = 2.9 \times 10^7 M_\odot$ (corresponding to $T_{vir} = 10^4 \rm K$) we expect 
a number of
BHs with mass between $M_1$ and $M_2$ (descendants of stars with mass between $m_1$ and $m_2$):
\begin{equation}
N_\bullet(M_1, M_2) = M_* \,  \nu_{SN} \left( \frac{\int_{m_1}^{m_2}  \phi(m) dm}{\int_8^{500} \phi(m) dm}\right),
\end{equation}
where $M_* = M_{vir} \left( {\Omega_b}/{\Omega_m} \right) f_* \approx 5\times 10^5 M_\odot$ for a star formation efficiency  $f_*=10 \%$.
We find $N_\bullet^{tot}\approx 1000$ BHs with mass between $2$ and $500$ $M_\odot$. Their mass distribution is shown in Table \ref{tab:pdf}:
the majority of the BH population settles between { $(M_1, M_2) = (2, 100) M_\odot$} and only $2\%$ is between $150$ and $500 M_\odot$.

\begin{table}
		\centering
		\begin{tabular}{c|c|c}
			\hline
			 $(m_1,m_2) [M_\odot]$                  & $(M_1, M_2) [M_\odot]$  & $N_\bullet/N_\bullet^{tot}$ \\
			\hline
			$25, 140$	& $2,100$ &$98\%$ \\
			$260, 500$	& $150,500$ &$2\%$ \\
			\hline
		\end{tabular}

\caption{ Mass distribution of the stellar black holes expected in a $T_{vir} = 10^4 \rm K$ halo at $z=10$. The $98\%$ sprouts in the $(2,100) M_\odot$ interval, while only $2\%$ is heavier than $100 M_\odot$. Stars are assumed to form with a Salpeter (1955) IMF and to leave a BH remnant according to Heger et al. (2002).}
\label{tab:pdf}
\end{table}	

Sampling the IMF and using the results from \cite{heger:2002}, we assign a mass to this population of 1000 BHs 
and generated the initial conditions of their motions in the galaxy $(r_0, v_0, \alpha)$ stochastically.
We followed their accretion rate $\dot{M}_\bullet(t)$ and integrated eq. \ref{eq:sync_spec} between $[0.5, 8] \rm \, keV$, obtaining their cumulative X-ray emission as a function of time (Fig. \ref{fig:lum}).
The X-ray luminosity flickers around a mean value of $\bar{L}_X = 6\times 10^{36} \rm erg s^{-1}$, with the spikes corresponding to 
passages through the densest parts of the disk.
We thus obtain the X-ray luminosity of stellar black holes per stellar unit mass formed:
\begin{equation}
\lambda_X = \bar{L}_X / M_* \approx 10^{31} \rm erg \, s^{-1} \, M_\odot^{-1}.
\end{equation}

To estimate
the stellar mass formed at a given time, we
assume that the Star Formation Rate (SFR) $\psi$ of the galaxy rises with time as
\begin{equation}
\psi (z)  = \psi_0 \left( 1- e^{-t(z)/\tau}\right),
\label{eq:sfr} 
\end{equation}
where $t(z)$ is the age of the universe at redshift $z$ with an unknown timescale $\tau > 0$ which parametrizes the uncertainty on the exact shape of $\psi(z)$. However, we set $\psi_0$ so that, for each value of $\tau$, the SFR is \citep{yue:2015} $10^{-3} M_\odot  \rm yr^{-1}$ at $z=10$.
It turns out that even allowing $\tau$ to vary in a very broad range, the total stellar mass in the galaxy is: 
\begin{equation}
M_* = \int^{10}_\infty \psi(z) dz \approx (2 - 5) \times 10^5 M_\odot, 
\label{eq:mstar}
\end{equation}
consistent with $f_* = \left(M_*/M_{vir}\right) \left( \Omega_m/\Omega_b\right) \approx 0.1$, previously assumed.

In Fig. \ref{fig:comparison} we show (red shaded region) the expected BHs $[0.5, 8] \rm \;  keV$ luminosity obtained as
\begin{equation}
	{L}_X(z) = \lambda_X M_*(z).
\end{equation}

It is interesting to compare such luminosity with the one expected from X-ray binaries ($L_{\rm XRB} $) in [0.5-8] keV in the same galaxy.
The local correlation with the SFR is \citep{mineo:2012, mesinger:2015}:
\begin{equation}
L_{\rm XRB}^0 = 3\times 10^{39} \rm \; erg \, s^{-1} \left( \frac{\psi}{ M_\odot \rm yr^{-1}}\right),
\label{eq:xrb}
\end{equation}
In general, high-$z$ galaxies are believed to show a lower metallicity which might lead to higher X-ray production efficiencies.
\cite{dijkstra:2012} \citep{lehmer:2016} found that the local law (eq. \ref{eq:xrb}) evolves up to $z=3$ ($z=7$) as follows:
\begin{equation}
L_{\rm XRB}(z) \approx L_{\rm XRB}^0 (1+z).
\label{eq:xrbz}
\end{equation}
In Fig. \ref{fig:comparison}, the XRB emission is shown with a blue line.

To conclude, we show the BH and XRB contribution to the galaxy SED at redshift $z=7$ (Fig. \ref{fig:cum_spec});
the red line represents the cumulative synchtrotron spectrum of the BH population predicted by eq. \eqref{eq:sync_spec};
the blue line corresponds to XRB.
According to \cite{rephaeli:1995}, \cite{swartz:2004}, \cite{mineo:2012}, the intrinsic XRB spectrum follows a power-law $\nu L_\nu \propto \nu^\alpha$, with $\alpha \approx 0.7-1.0$ and a cut-off above 10 keV \citep{miyawaki:2009, bachetti:2013}. Following \cite{pacucci:2014} we take $\alpha = 0.8$ as the fiducial value; then we normalized the SED so that it reproduces $L_{\rm XRB} (z=7)$ predicted by eq. \eqref{eq:xrbz} with $\psi = 10^{-3} M_\odot /\rm yr$.

Figures \ref{fig:comparison} and \ref{fig:cum_spec} lead to the conclusion that 
X-ray emission from stellar mass BH is comparable to that of XRB.
Thus, the result of this study suggests that the stellar BH energy input should be included in IGM heating computations
relevant to the HI 21 cm line signal from cosmic dawn.

\begin{figure}
                \centering
                \includegraphics[width=0.5\textwidth]{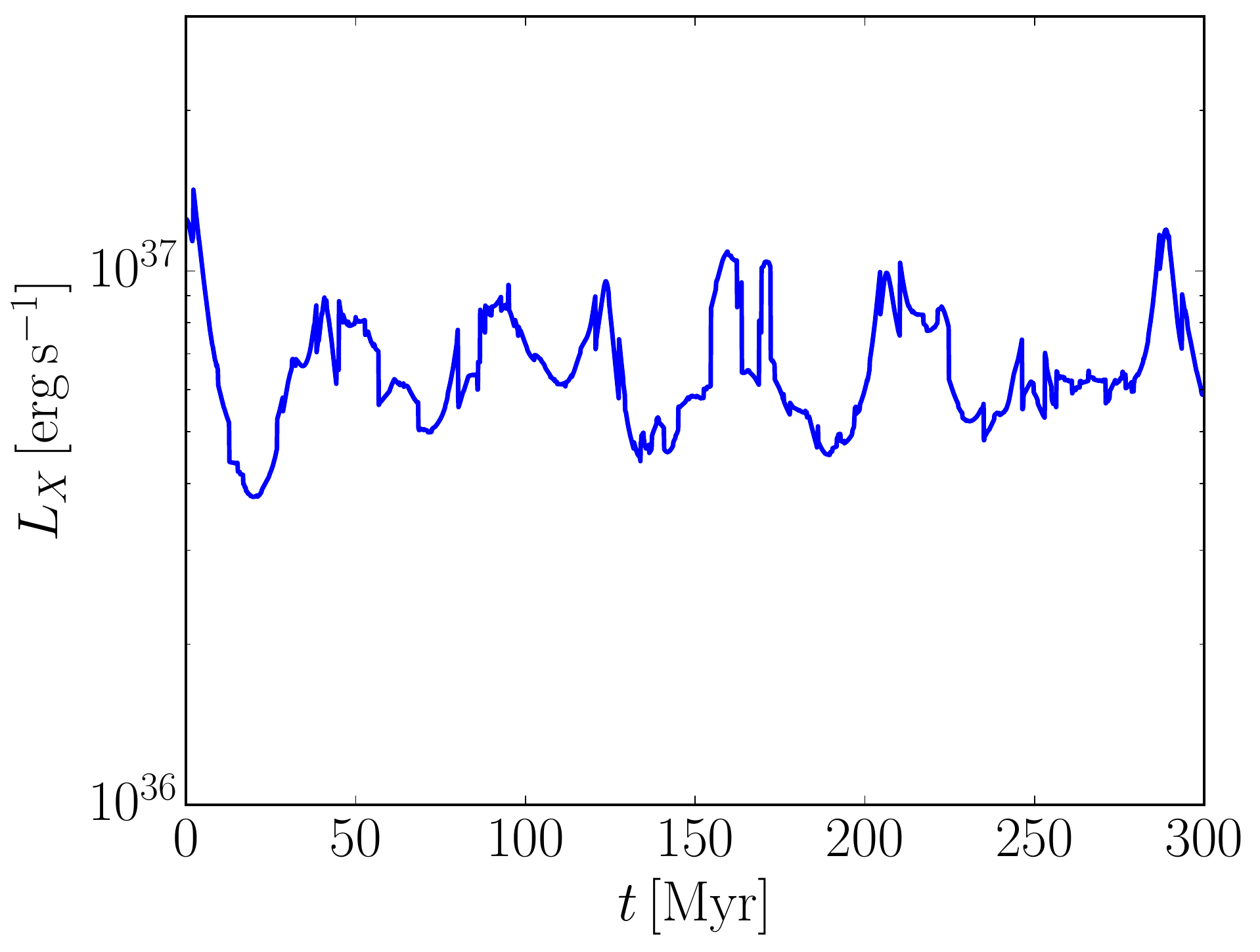}
                \caption{Cumulative $[0.5-8] \, \rm keV$ emission of a 1000 stellar black holes (sBH) sample hosted in the modelled galaxy as a function of time: it flickers around a mean value of $\bar{L}_X = 6\times 10^{36} \rm erg s^{-1}$, with the spikes corresponding to passages through the densest parts of the galactic disk.}
                \label{fig:lum}
\end{figure}

\begin{figure}
                \centering
                \includegraphics[width=0.5\textwidth]{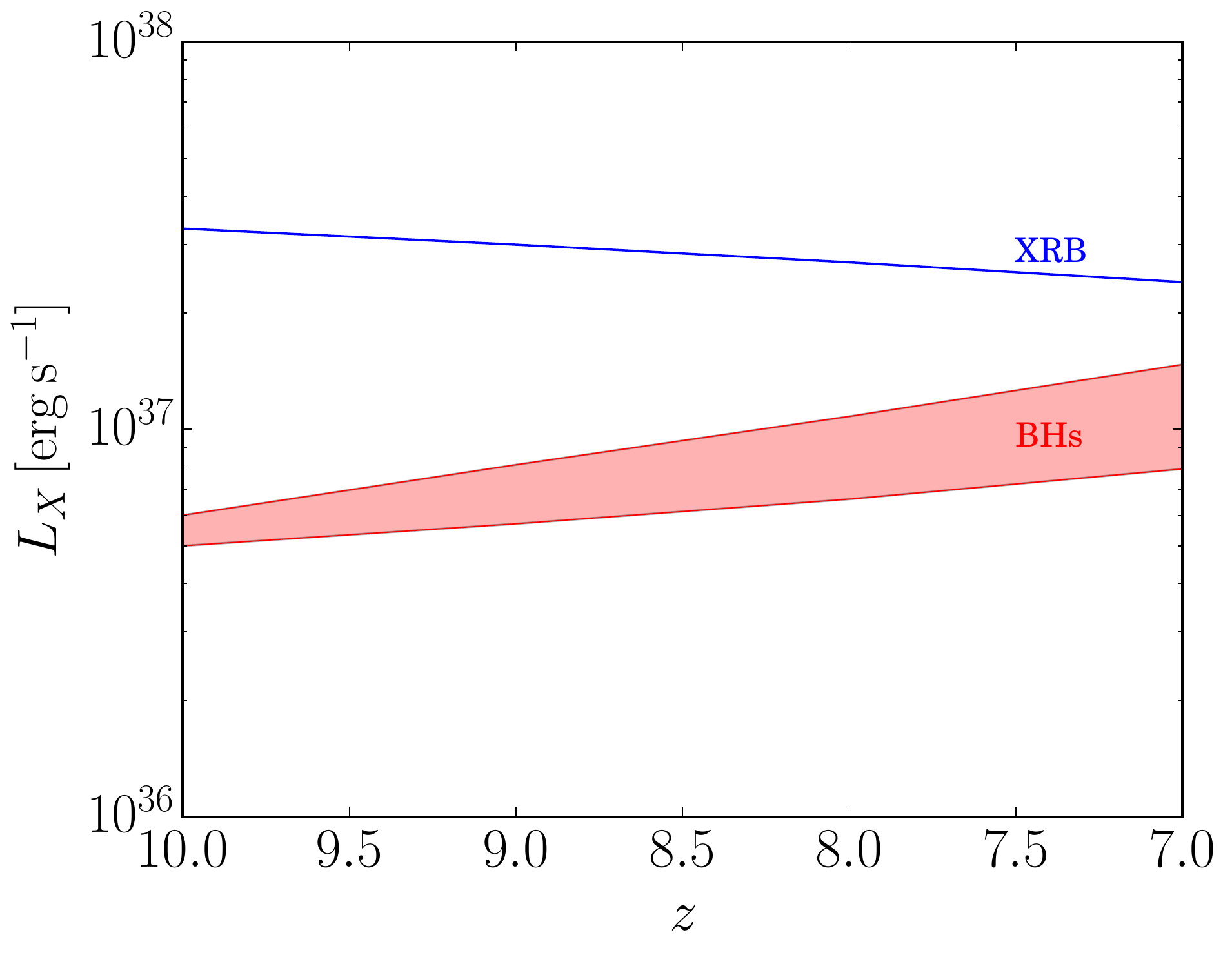}
                \caption{[0.5-8] keV emission of the stellar BHs population (red shaded region), compared to the one of the XRB (blue line), as a function of redshift. XRB contribution at high-$z$ relies on the polynomial law eq. 42 extrapolated up to $z=10$. The error on BHs emission highlights the uncertainty on the exact functional shape of the SFR $\psi(z)$.}
                \label{fig:comparison}
\end{figure}

\begin{figure}
                \centering
                \includegraphics[width=0.5\textwidth]{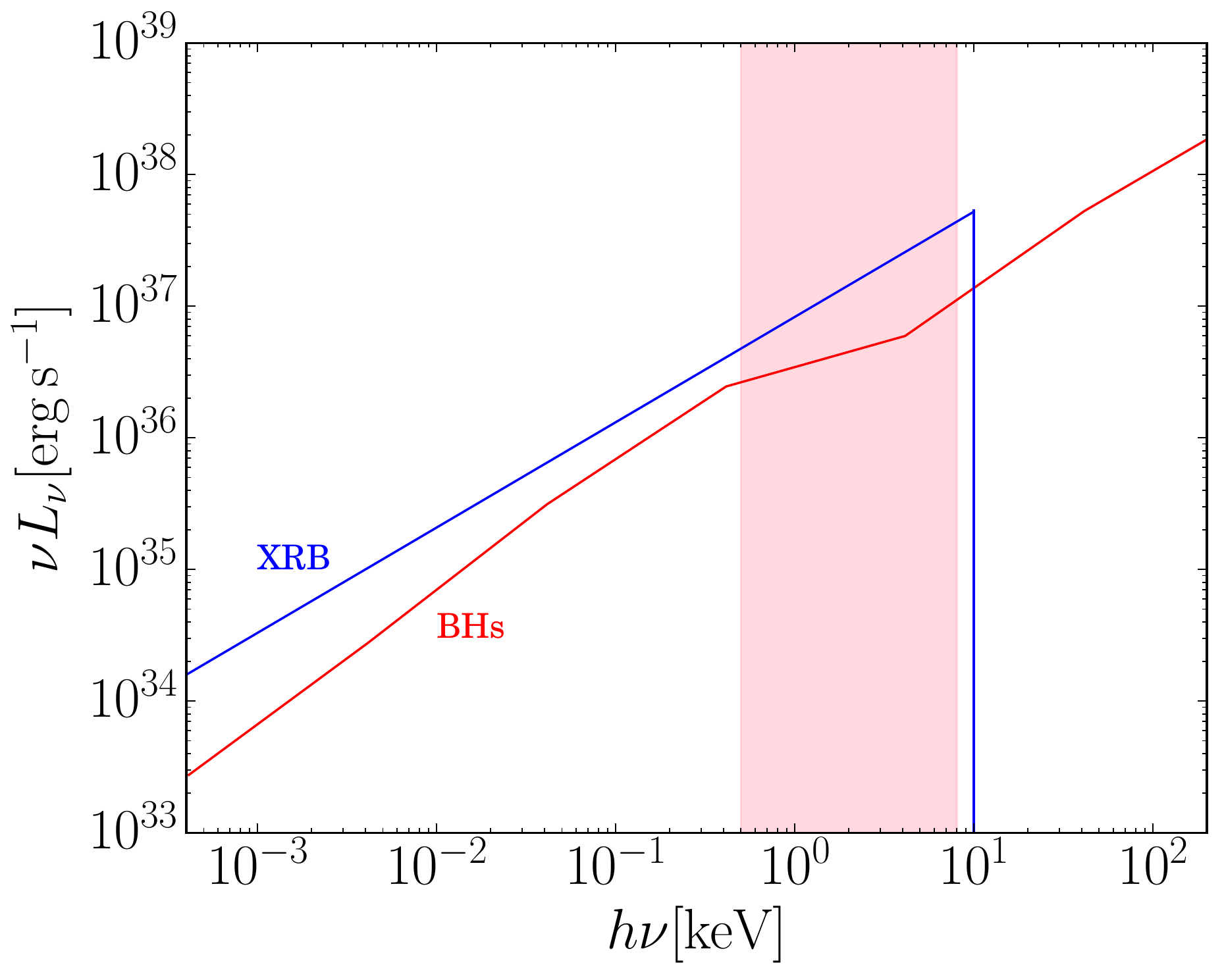}
                \caption{Comparison between the BHs cumulative SED (eq. \ref{eq:sync_spec}) and the XRB SED \citep{pacucci:2014} at z=7. It is noteworthy that our model predicts BHs to emit above $10 \, \rm keV$, where XRB spctrum is indeed observed to crash. The pink shaded region corresponds to the energy interval [0.5-8] keV.}
                \label{fig:cum_spec}
\end{figure}

\section{Summary and discussion}
\label{sec:conclusions}
We studied the accretion rate of a $M_\bullet = 100 M_\odot$ BH formed in $z=10$ galaxies under different conditions (e.g. galaxy  structure, BH initial position and velocity). We modelled the galaxy baryonic content and followed the BH orbit and accretion history for $300 \, \rm Myr$ { (time interval between $z=7$ and $z=10$)}, assuming the radiation-regulated accretion model by \cite{park_ricotti:2013}.
We analyzed the mass increment $\Delta M$ of the BH in the integration time as a function of the orbital parameters. 

We found that, within the limits of our model, the black hole cannot grow by more than $30 \%$; this suggests that simple accretion on light-seed models are inadequate to explain high-$z$ SMBH, and strongly encourages further studies on heavier seeds candidates and/or the occurrence of merging events.

We finally modelled the [0.5-8] keV emission from such stellar BH population, providing an estimate of its cumulative luminosity in this band.
We further compute the expected emission from XRB in the same band, concluding that the one from stellar mass BH is lower than, but still comparable, with the one from XRB.
Thus, the result of this study suggests that the stellar BH energy input should be included in IGM heating computations
relevant to the HI 21 cm line signal from cosmic dawn. A detailed prediction of the effects of such X-ray emission on the 21 cm line power spectrum requires a future dedicated study.

Our simple model can be improved in many ways by including, e.g. disk formation around the BH, different (realistic) gas density distribution and effects of dynamical friction on BH's orbit, but it seems that none of them are likely to change 
the conclusion we have drawn in a substantial manner. 

Concerning the formation of an accretion disk, we considered the rapid BH growth under anisotropic radiation feedback by \cite{sugimura:2017}; it hypothesizes the BH to sprout and settle in the center of the galaxy and grow at $55\% \dot{M}_B$ leading, in our case to a BH of $122 M_\odot$.
Almost the same mass increment results by accounting for the presence of denser molecular clouds. 
\cite{pallottini:2017}, performed  simulations of high-$z$ galaxies to characterize their internal structure, finding $\rm H_2$ concentrated in a disk-like structure that is $0.01 \%$ of the galaxy total volume and has $n \approx 25 \, \rm cm^{-3}$; this density does not lead to significant difference with our former results.
In any case, even if this density would rise up to $10^5 \rm cm^{-3}$ \citep{solomon:1987, bolatto:2008, mckee:ostriker:2007}, $\Delta M_{max} \approx 23 M_\odot$.
Since the nearer to the galaxy center the denser is the gas, the effect of dynamical friction on the orbit produces a slight increase of the accretion rate: 
we adopted the model by \cite{tagawa:2016} for the acceleration due to the dynamical friction on the BH and found out a double  $\Delta M_{max}$ with respect to our naiver formulation.

To conclude, even assuming the stellar mass BHs to accrete at the Eddington limit, they could grow up to $1 \, (2)$ order of magnitude if the radiative efficiency is $0.1 \, (0.01)$;
furthermore, applying our Eddington limited accretion model to $10^{4} - 10^6 M_\odot$ BHs
we obtain a mass increase of at most $30 \%$ of their initial mass.

The solution to the problem of high-$z$ quasars seems  then to require high mass seeds and merging/super-Eddington accretion episodes.
For what concerns merging, the results of the zoom-in cosmological hydrodynamical simulation presented by \cite{barai:2018} suggest that seeds growth is indeed dominated by direct accretion.  
For what concerns super-Eddington accretion, several authors have shown that it requires stringent and specific conditions, as
for example slim accretion-disks \citep{madau:2014, volonteri:2015} or seeds to be anchored in a dense,  gas-rich star cluster \citep{alexander:2014}. 

\bibliographystyle{mnras}
\bibliography{biblio.bib}
\bsp

\label{lastpage}
\end{document}